\documentclass[pdflatex,sn-mathphys-ay]{sn-jnl}% Math and Physical Sciences Numbered Reference Style
\usepackage{graphicx}%
\usepackage{multirow}%
\usepackage{amsmath,amssymb,amsfonts}%
\usepackage{amsthm}%
\usepackage{mathrsfs}%
\usepackage[title]{appendix}%
\usepackage{xcolor}%
\usepackage{textcomp}%
\usepackage{manyfoot}%
\usepackage{booktabs}%
\usepackage{algorithm}%
\usepackage{algorithmicx}%
\usepackage{algpseudocode}%
\usepackage{listings}%
\usepackage{setspace}

\theoremstyle{thmstyleone}%
\theoremstyle{thmstyletwo}%

\theoremstyle{thmstylethree}%
\begin{document}

\title[The ELFIN EPD]{The Energetic Particle Detector Instrument on ELFIN}

%%=============================================================%%
%% GivenName	-> \fnm{Joergen W.}
%% Particle	-> \spfx{van der} -> surname prefix
%% FamilyName	-> \sur{Ploeg}
%% Suffix	-> \sfx{IV}
%% \author*[1,2]{\fnm{Joergen W.} \spfx{van der} \sur{Ploeg} 
%%  \sfx{IV}}\email{iauthor@gmail.com}
%%=============================================================%%

\author*[1]{\fnm{Colin} \sur{Wilkins}}\email{colinwilkins@ucla.edu}

\author[1]{\fnm{Ryan} \sur{Caron}}\email{rcaron@epss.ucla.edu}

\author[1]{\fnm{Ethan} \sur{Tsai}}\email{ethantsai@ucla.edu}

\author[1]{\fnm{Chris} \sur{Shaffer}}\email{chrisshaffer@ucla.edu}

\author[7,1]{\fnm{Patrick} \sur{Cruce}}\email{pcruce@gmail.com}

\author[12]{\fnm{Maxwell} \sur{Chung}}\email{maxwellchung.uc@gmail.com}

\author[1]{\fnm{Andrei} \sur{Runov}}\email{andrei@ucla.edu}

\author[1]{\fnm{Jiang} \sur{Liu}}\email{jliu@igpp.ucla.edu}

\author[3]{\fnm{Drew} \sur{Turner}}\email{Drew.Turner@jhuapl.edu}

\author[11,1]{\fnm{Yuri} \sur{Shprits}}\email{yshprits@gfz.de}

\author[6]{\fnm{Wen} \sur{Li}}\email{wenli77@bu.edu}

\author[4]{\fnm{Michael} \sur{Lawson}}\email{mlawson@blueorigin.com}

\author[1]{\fnm{Emmanuel} \sur{Masongsong}}\email{emasongsong@epss.ucla.edu}

\author[1]{\fnm{Cindy} \sur{Russell}}\email{cindy.russell@rocketmail.com}

\author[5]{\fnm{Eric} \sur{Grimes}}\email{eric.grimes@nasa.gov}

\author[7]{\fnm{Jiashu} \sur{Wu}}\email{jiashu8@ualberta.ca}

% students

\author[1]{\fnm{Matthew} \sur{Allen}}\email{matt.allen@flyzipline.com}

\author[14]{\fnm{Jessica} \sur{Artinger}}\email{jessica.artinger@gmail.com}

\author[10]{\fnm{Alex} \sur{Flemming}}\email{flemming.adf@gmail.com}

\author[1]{\fnm{Brayden} \sur{Hesford}}\email{Braydenhesford@gmail.com}

\author[1]{\fnm{Laura} \sur{Iglesias}}\email{lauraiglesias4@gmail.com}

\author[1]{\fnm{James} \sur{King}}\email{jcking1034@gmail.com}

\author[1]{\fnm{Austin} \sur{Norris}}\email{norris.austinc@gmail.com}

\author[1]{\fnm{Carter} \sur{Pedersen}}\email{carter.e.pedersen@gmail.com}

\author[3]{\fnm{Daniel} \sur{Thomlinson}}\email{daniel.thomlinson@aero.org}

\author[9]{\fnm{Wynne} \sur{Turner}}\email{wynnet@ucr.edu}

\author[1]{\fnm{Vassilis} \sur{Angelopoulos}}\email{vassilis@ucla.edu}

\affil*[1]{\orgdiv{Earth, Planetary, and Space Sciences}, \orgname{University of California, Los Angeles}, \orgaddress{\street{595 Charles E. Young Dr. East}, \city{Los Angeles}, \postcode{90095}, \state{CA}, \country{USA}}}
%\affil[2]{TBD pending email response}
\affil[2]{The Aerospace Corporation, El Segundo, CA, USA}
\affil[3]{Applied Physics Laboratory, Johns Hopkins University, North Laurel, MD, USA}
\affil[4]{Blue Origin, Los Angeles, CA, USA}
\affil[5]{NASA Goddard Space Flight Center, Greenbelt, MD, USA}
\affil[6]{Center for Space Physics, Boston University, Boston, MA, USA}
\affil[7]{University of Alberta, Alberta, Canada}
\affil[8]{Apple Computer, Cupertino, CA, USA}
\affil[9]{University of California, Riverside, Riverside, CA, USA}
\affil[10]{PsiQuantum, Palo Alto, CA, USA}
\affil[11]{GFZ Potsdam, Potsdam, Germany}
\affil[12]{Northrop Grumman, Los Angeles, CA, USA}
\affil[13]{Orange Coast College, Costa Mesa, CA, USA}
%\affil[]{}

%\affil[2]{\orgdiv{Department}, \orgname{Organization}, \orgaddress{\street{Street}, \city{City}, \postcode{10587}, \state{State}, \country{Country}}}

%\affil[3]{\orgdiv{Department}, \orgname{Organization}, \orgaddress{\street{Street}, \city{City}, \postcode{610101}, \state{State}, \country{Country}}}

%%==================================%%
%% Sample for unstructured abstract %%
%%==================================%%

\abstract{We present the design methodology, implementation, and data products of the Energetic Particle Detector (EPD) instrument on-board the Electron Losses and Fields Investigation (ELFIN) dual SmallSat mission. ELFIN was designed to study the time-evolution of energetic particle distributions in Earth's magnetosphere, having operated in polar Low Earth Orbit (LEO) from 2018 to 2022. Each satellite's instrument payload contained a pair of detector heads to sense differential-directional fluxes of 50 keV to $\sim$5 MeV electrons (EPDE), and 50 keV to $\sim$2 MeV atomic ions (EPDI), with an energy resolution better than 50\% ($\Delta E/E<0.5$). The satellites were spinning with a nominal period of 2.8 s in a plane containing Earth's background magnetic field, with particle counting intervals subdivided into 16 consecutive spin sectors. Each EPD was paired with a three-axis Fluxgate Magnetometer (FGM) sampled at 80 Hz for local pitch-angle determination. Each detector head operated on the principle of pulse-height analysis (PHA) in solid state media, collimated by a rectangular 22.5$^\circ$ field of view and entrance aperture. The EPDE contained a stack of 6 silicon detectors (0.7 cm total thickness) with 0.15 cm$^2$ str geometric factor, with ion contamination mitigated by an aperture foil and digital coincidence logic. The EPDI contained a stack of 2 silicon detectors (0.1 cm total thickness) with 0.03 cm$^2$ str geometric factor, with electron contamination mitigated by a magnetic broom stage and digital anti-coincidence logic. The total assembled mass, volume, and power required for the EPD detector heads and circuit boards were 939 g, 10$\times$9.8$\times$8.4 cm$^3$, and 2.3 W, respectively. We describe each EPD design element alongside instrument data products and processing.}


%%\pacs[JEL Classification]{D8, H51}

%%\pacs[MSC Classification]{35A01, 65L10, 65L12, 65L20, 65L70}

\maketitle

\section{Introduction}
The Electron Losses and Fields Investigation with a Spatio-Temporal Ambiguity-Resolving option (ELFIN) was a pair of identical 3-Unit (3U) CubeSats in a nearly circular, polar ($\sim$93$^\circ$ inclination), low-Earth ($\sim$450 km) orbit, launched together on 15 September 2018 \citep{Angelopoulos2020}. Its primary objective was to resolve the mechanism of storm-time relativistic electron precipitation, for which pitch-angle scattering by electromagnetic ion cyclotron (EMIC) waves is a leading candidate. From low altitude, ELFIN's pitch-angle-resolving measurements test whether relativistic electron spectra within the bounce loss cone carry the distinct signatures of EMIC scattering rather than of chorus scattering or magnetopause shadowing. Pairing two satellites at a slowly varying along-track separation helps to distinguish spatial structure from temporal evolution on timescales faster than a single satellite's $\sim$90 min orbit. ELFIN was operated to further capitalize on the presence of the equatorial Van Allen Probes, Arase (ERG), and THEMIS missions, which measure the waves and near-equatorial populations that drive precipitation, extending low-altitude spacecraft conjunction studies with these missions in the energy range, pitch-angle and spatial distribution of chorus-driven precipitation and of net precipitation and backscatter rates from 50 keV to several MeV \citep{Angelopoulos2020}. The measurements underpinning these objectives are made by the Energetic Particle Detector (EPD), the key instrument of the payload. 

EPD comprises two solid-state detector heads (alternatively, ``sensor heads''): an electron head (EPDE) covering 50 keV–5 MeV, and an ion head (EPDI) covering $\sim$50 keV to several MeV, each resolving 16 energy channels at $\Delta$E/E $<$ 50\% \citep{Angelopoulos2020}. With the spin axis normal to the orbit plane and a $\sim$3 s spin, EPD sweeps the full range of local pitch angles twice per spin, sampling loss-cone and trapped populations on every rotation. The fluxgate magnetometer (FGM) supplies, in flight, the spin-plane field phase that EPD uses to sort counts into pitch-angle sectors on board, allowing energy- and spin phase-resolved spectrograms to be produced without post-facto coordinate transformations or FGM calibration. This combination — a pitch-angle-resolving energetic-particle spectrometer flown with a magnetometer on a spinning, low-altitude, polar platform — gives ELFIN capabilities no previous mission possessed. 

The EPD was described only briefly in the mission overview paper \citep{Angelopoulos2020}. This dedicated instrument paper (1) expands the description of the EPD sensor, electronics, and operating principle; (2) demonstrates, after in-orbit calibration, that EPDI performs to its 50 keV design floor on the nightside — in shadow, free of direct sunlight and Earth-glow — correcting the mission paper's report that the ion head was off-nominal, and yielding roughly four months of clean ion data; (3) reports the mature calibration and analysis tools behind the Level-1 (L1) and Level-2 (L2) data products; and (4) illustrates compound scientific examples enabled by EPDE, EPDI, and FGM operating together. The instrument requirements detailed later follow directly from these mission objectives, which we first introduce by establishing EPD scientific traceability and mission heritage.

\subsection{EPD Requirements and Adherence}
ELFIN's objectives translate into a compact set of measurement requirements concerning energy resolution, time resolution, angular resolution, and minimum detectable flux (sensitivity). The minimum (primary) objective demands a unidirectional electron detector that resolves loss-cone spectra from a spinning platform, with pitch-angle resolution better than $\sim$30$^\circ$ and $\Delta$E/E $\sim$ 50\%, paired with an attitude magnetometer \citep{Angelopoulos2020}. Attitude
knowledge is apportioned between magnetic-field knowledge ($<$2$^\circ$) and particle field of view ($<$28$^\circ$). The secondary objectives add only modest capability: the EPDI ion head and extension of the electron range down to $\sim$0.1 MeV, both inherited from ELFIN-L. The as-built EPD meets or exceeds these requirements. EPDE resolves pitch angle to $\pm$11.25$^\circ$ (FWHM $<$ 22.5$^\circ$), so at $\sim$400–450 km (where the $L = 4$ equatorial loss cone maps to local pitch angles of $\sim$65$^\circ$) several sectors sample the loss cone, while at least one sector near 90$^\circ$ samples the trapped population. Each radiation-belt crossing (L $\approx$ 2.5–7) is captured with 16 sectors per spin and 16 energies per species, at least once per four orbits. Each selected crossing provides spin-resolved measurements, with a baseline observing cadence of at least one crossing every four orbits. This baseline time resolution is comparable to the rate at which the storm-time belt evolves, and can be reconfigured by command to use up to 32 sectors for increased fidelity during interesting geomagnetic intervals. EPDI's broom-magnet system (0.28 T gap field) sweeps electrons up to 500 keV out of the ion aperture. This allows the geometric factor to be maximized while avoiding saturation for 50 keV ions at active times. 

The mission paper reported EPDI exhibited off-nominal performance, quoting a $\sim$250 keV low-energy floor attributed to the ion detectors' inactive (dead) layer. However, later in-flight calibration, including re-tuning of detector anti-coincidence logic and thresholds, showed this conclusion to be incorrect. The issue stemmed from the fact that optical contamination in the EPDI was present when performing the detector noise threshold procedure, leading to overly aggressive flux attenuation at low energies. Later in the mission, this procedure was properly performed in shadow, leading to the recovery of the original target energy range. The EPDE did not experience this problem due to the presence of an aperture foil to block sunlight. For the EPDI, When the satellites are on the dayside, dayglow indeed gives rise to photon fluxes that the detector's nominal $\sim$1000 Å aluminum flashing cannot suppress. However, on the nightside, in shadow, ion measurements are clean down to 50 keV as intended. The roughly four months of good nightside ion data represent a previously unreported instrumental result, while the data was analyzed in works such as \cite{Wilkins2025}.

Table \ref{tab:L1_reqs} recasts ELFIN's L1 mission requirements into the derived EPDE and EPDI instrument requirements; Table \ref{tab:inst_performance} (from Table 3 of \cite{Angelopoulos2020}, with the EPDI energy-range entry updated for the calibration reported here) gives the demonstrated performance.

\subsection{EPD Historical Context and Heritage}
With one exception, no earlier polar-orbiting mission measured energetic electron and ion distributions at high latitude from low-Earth orbit with simultaneously high pitch-angle resolution, energy resolution, and cadence. The exception is the particle instrument on the sun-synchronous, noon–midnight satellite 1972-076B (98.4$^\circ$ inclination, $\sim$740 km), which, like ELFIN, spun about an axis perpendicular to its orbit plane \citep{Imhof1979}. Its 90$^\circ$-mounted electron spectrometer swept all pitch angles and cleanly measured precipitating and trapped fluxes over 0.16–2.4 MeV \citep{Imhof1979}. Those data defined the energy-dependent onset of pitch-angle isotropy near the trapping boundary — the isotropy boundary (IB) — with the isotropy threshold energy falling steeply with increasing L \citep{Imhof1977}, later refined during disturbed times \citep{Imhof1979}. The work was not sustained, largely because the datasets did not remain available. Electron spectrometers on UARS later extended isotropy-threshold measurements over 30 keV–5 MeV and established the IB as a remote-sensing diagnostic of the field configuration near the outer belt edge \citep{Imhof1997}, though UARS's $\sim$57$^\circ$ inclination reached only moderate latitudes and encountered the auroral oval only occasionally. 

Operational NOAA/TIROS and POES instruments illustrate the non-spinning limitation. Their Medium-Energy Proton and Electron Detector (MEPED) uses two fixed telescope pairs — one along the zenith (nominally loss cone), one at $\sim$80-90$^\circ$ (nominally trapped) — on a three-axis-stabilized platform, with integral electron channels ($>$30, $>$100, $>$300 keV) \citep{Hill85,Rodger2010}. Because the angle between the local field and the fixed axes changes along the orbit, the narrow fields of view cross-contaminate, and integral channels give little spectral detail. A spinning platform like ELFIN's, whose single detector sweeps all pitch angles relative to the measured field, avoids this by construction. POES/SEM-2 zenith telescopes additionally suffer severe proton contamination ($>$40\%, up to 55\% when disturbed) in the heart of the belts (L $\approx$ 4.5–5.5), complicating loss-rate estimates \citep{Rodger2010}, which is mitigated on ELFIN by the shielding design and detector coincidence logic.

SAMPEX delivered key low-altitude precipitation science from a complementary design. Its HILT sensor combined a very large geometric factor with 10 Hz sampling and zenith pointing \citep{Blake1996}, establishing the morphology of relativistic-electron precipitation: persistent latitudinal bands near the outer-belt edge and sub-second microbursts \citep{Nakamura1995}, the microbursts concentrated at L $\approx$ 4–6 and in the dawn sector during early storm recovery, pointing to whistler-mode scattering \citep{Nakamura2000}. HILT's angular information came only from the relative rates of its four detector rows, which internal scattering smeared so badly that they could establish only whether the loss cone was filled, not a pitch-angle spectrum \citep{Blake1996}. Its coarse, effectively integral energy response left the driver of these bands ambiguous: revisiting the SAMPEX statistics with conjugate POES and pitch-angle- and energy-resolved ELFIN data, \cite{Sun2026} show that the $>$1 MeV nightside bands can arise from field-line curvature scattering, EMIC waves, or whistler-mode waves, and so have no single driver — the ambiguity that only pitch-angle- and energy-resolved low-altitude measurements resolve. ELFIN's spin-resolved sampling of loss-cone and trapped fluxes across 16 energies removes it. 

The same isotropy phenomenon was established at the equator: on OGO 5, a scanning electron spectrometer and vector magnetometer used the isotropic-to-butterfly pitch-angle transition to trace the near-tail field versus radial distance \citep{West1978a,West1978b}. Equatorial spacecraft can therefore also probe field-line scattering, but their long orbital periods make it hard to obtain the near-instantaneous latitudinal profiles a $\sim$100 min polar orbit yields, against which the tail configuration can be tracked as it evolves \citep{Imhof1979}. This motivates renewed, modern IB studies from a capable low-altitude platform — a role EPD is well suited to fill. 

Low-altitude energetic-particle detectors locate magnetospheric boundaries and place auroral features relative to them, especially alongside ground-based imaging. The IB is the low-altitude signature of the transition between adiabatic and chaotic (current-sheet-scattered) motion; because that transition is set by the equatorial field, the IB position remotely senses the near-tail configuration, which the isotropic-boundary algorithm exploits to reconstruct the equatorial field \citep{Sergeev1993}. The ion energy-flux peak (b2i) tracks the proton IB and indexes magnetotail stretching, though spacecraft such as DMSP cannot determine the IB directly, measuring only precipitating fluxes \cite{Newell1998}. Combining low-altitude profiles with THEMIS all-sky imagers and in-situ data, \cite{Sergeev2012} showed the equatorward prebreakup arc coincides with the energy-dispersed electron IB and its energetic-electron arc — an analysis requiring exactly the pitch-angle-resolved, ground-conjugate measurements EPD provides.

The EPD detector derives from the Solid State Telescope (SST) on the five-satellite THEMIS mission, sharing its front-end processing with mass and volume adapted for the CubeSat \citep{Angelopoulos2008}. THEMIS spans the near-equatorial magnetotail, magnetopause, and solar wind, and its SST was itself built to the heritage of similar solid-state telescopes on Wind and STEREO. Wind's 3DP double-ended telescopes separate species with a thin foil on one end (counting electrons) and a broom magnet on the other (counting ions) \citep{Lin1995}; STEREO carries the analogous IMPACT suprathermal-electron and solid-state detectors \citep{Luhmann2007,Lin2008}, and MAVEN's SEP sensors are closely based on the THEMIS SST \citep{Larson2015}. These motivated two features of EPD in particular — the solid-state telescope and the broom-magnet species separation (in MAVEN SEP a $\sim$0.25 T field sweeps electrons below 350 keV) \citep{Larson2015}. The connection is one of operating principle, not geometry: the Wind/THEMIS/STEREO/MAVEN designs measure electrons and ions back-to-back at opposite ends of one double-ended telescope, whereas ELFIN separates them into dedicated EPDE and EPDI heads \citep{Angelopoulos2020}. EPD's shielding and anti-coincidence design also draw on the OGO 5 lineage through the UCLA particle-detector experience \citep{Angelopoulos2020}. 

ELFIN's instrument was first built and flown as ELFIN-L, the zenith-only suite on the non-spinning Lomonosov spacecraft \citep{Shprits2017}. Lacking spin, ELFIN-L had a single, roughly zenith-pointing look direction and could not resolve pitch angle, measuring the precipitating flux (and the locally trapped flux where the orbit geometry allowed, as over the South Atlantic Anomaly) from a 400–500 km sun-synchronous orbit. Its purpose was to determine precipitating electron and ion spectra and, above all, to prove the instrument in flight ahead of the spinning CubeSats \citep{Shprits2017}. Delivered in July 2011, it returned the first EPD electron-precipitation spectra, without pitch-angle information \citep{Angelopoulos2020}. ELFIN is the maturation of this line: a THEMIS-derived solid-state telescope, first proven in polar orbit on ELFIN-L, now flown with full pitch-angle resolution — resolving both precipitating and trapped populations — and a companion magnetometer on two spinning CubeSats. 

\begin{figure}[!htb]
    \centering
    \includegraphics[width=0.5\linewidth]{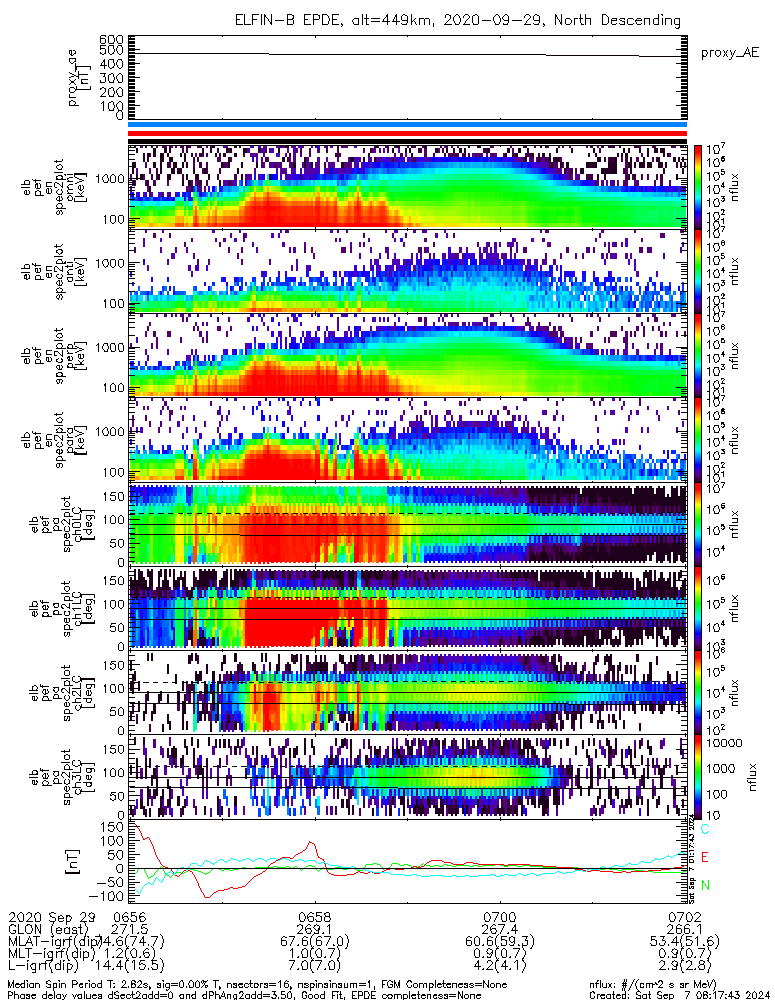}\includegraphics[width=0.5\linewidth]{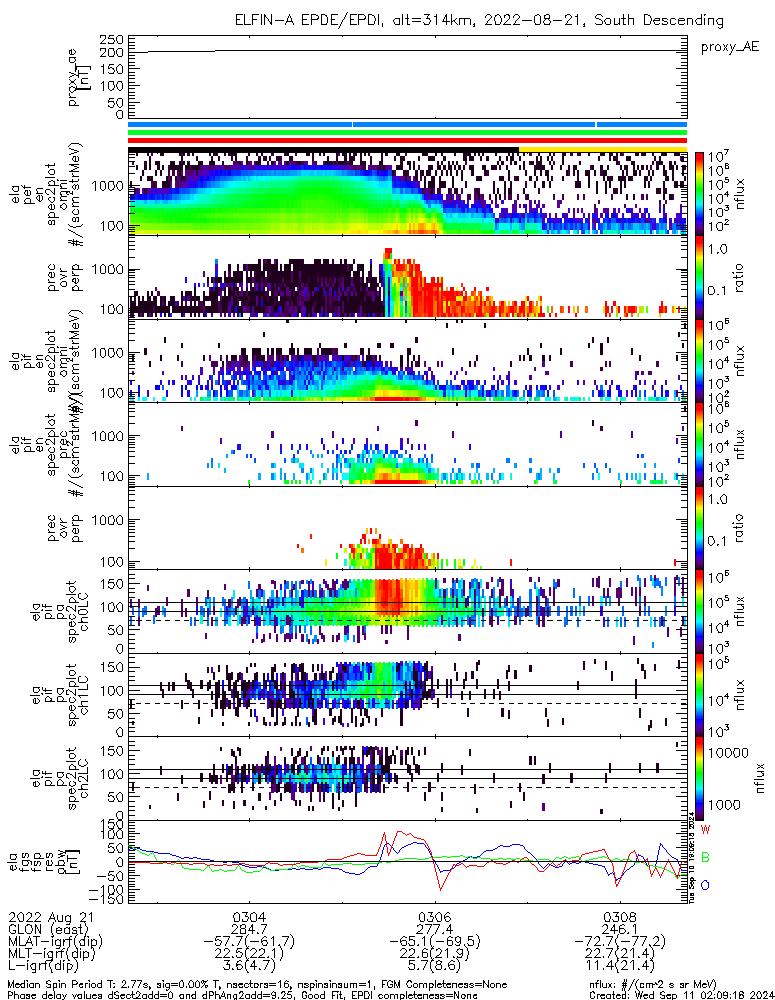}
    \caption{Example summary plots containing ELFIN EPD Level-2 electron energy- and pitch-angle spectra (EPDE; left) collected by ELFIN-B in a North Descending pass on 2020-09-29 from 0656-0702 UT, and combined electron and ion energy- and pitch-angle spectra (EPDE+EPDI; right) collected in a South Descending pass by ELFIN-A on 2022-08-21 from 0303 to 0309 UT. Descriptions and processing methods corresponding to each panel are presented in the Data Products section.}
    \label{fig:overview_plots}
\end{figure}

As a preview of the ELFIN EPD dataset, two examples of science collections, taken from the mission summary plots, are shown in Fig. \ref{fig:overview_plots}. The left panels show electron number flux spectrograms versus energy and pitch-angle over a typical 6-7 minute collection interval. The right panels demonstrate a combined electron-ion detection --- an EPD capability that had not been previously reported. Specifics of these panels, as well the underlying data processing, are described in Section 6.

The remainder of this manuscript is outlined as follows: Section 2 presents the EPD principles of operation, including a brief summary of particle detection in solid-state media. Section 3 details the detectors and mechanical design. Section 4 covers the electrical and digital system design. Section 5 covers instrument calibration and commissioning, including particle tracing and radioactive source testing. Section 6 covers EPD data acquisition and products, including magnetic spin sectoring, the L1 and L2 data processing, the tools for accessing them, and examples of science enabled by EPDE, EPDI, and FGM together.

\section{Instrument operational principles}
\begin{figure}[!htb]
    \centering
    \includegraphics[width=\linewidth]{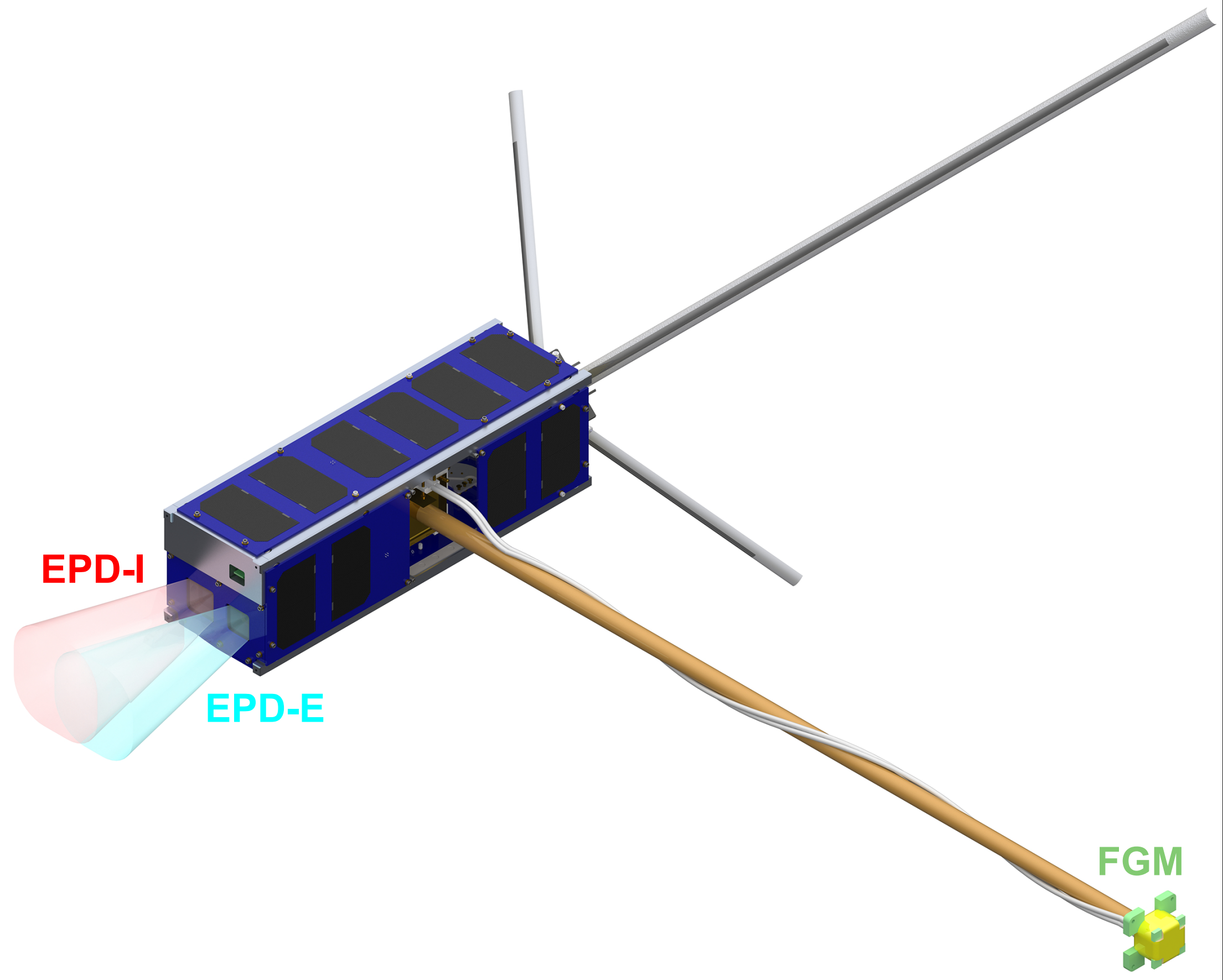}
    \caption{View of EPD electron (EPDE) and ion (EPDI) detector heads in relation to spacecraft and Fluxgate Magnetometer (FGM), held on a deployable boom. The satellites nominally spin perpendicular to the EPD look directions and FGM axis in a plane containing the background magnetic field for particle pitch-angle determination.}
    \label{fig:epd_fovs}
\end{figure}

\begin{figure}[!htb]
    \centering
    \includegraphics[width=\linewidth]{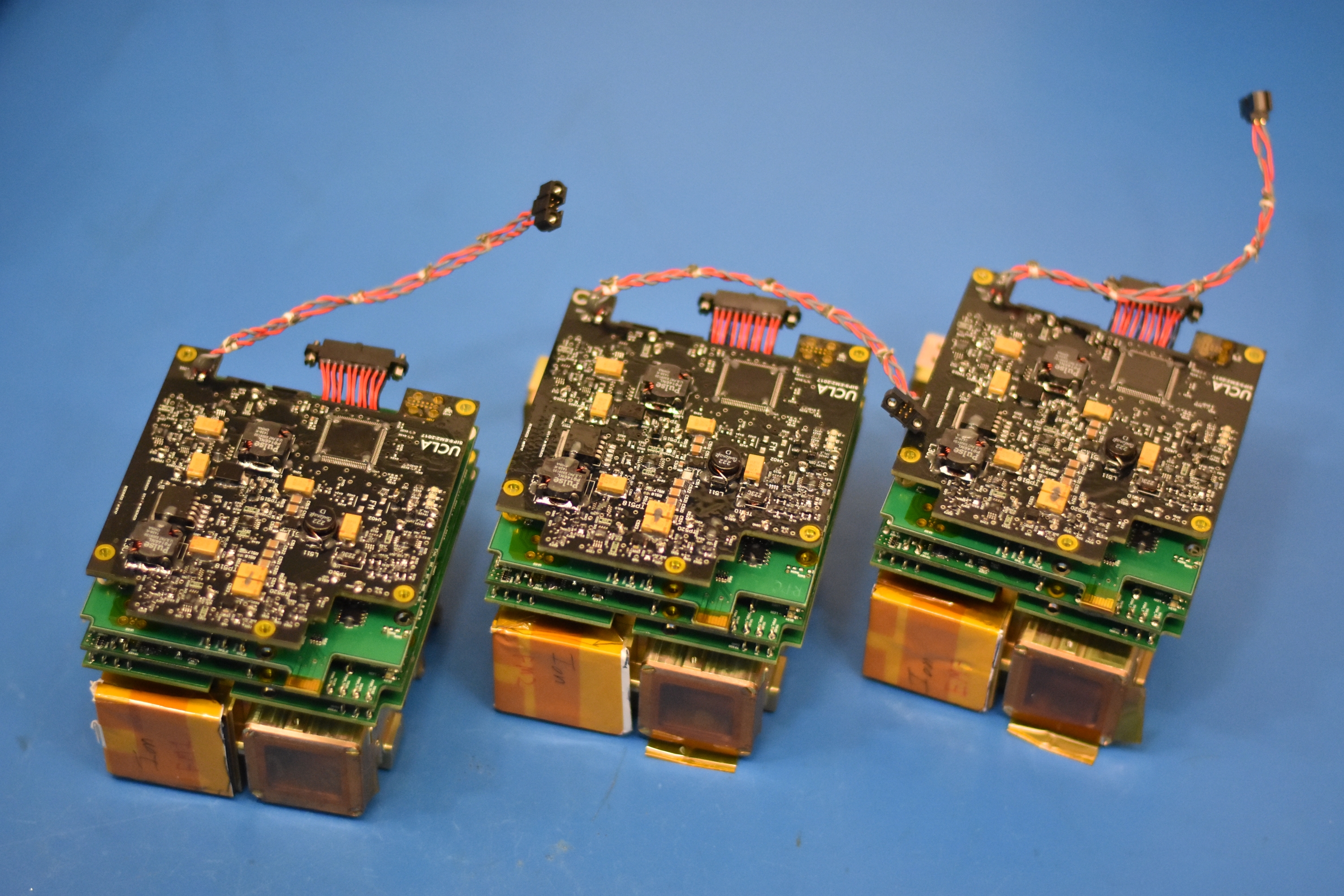}
    \caption{External view of the assembled EPD/detector heads, including Flight Model 1 flown on ELFIN-A (FM-1; left), Flight Model 2 on ELFIN-B (FM-2; middle), and Engineering Model 3 (EM-3; right) used as a ground reference. The electronics board stack and interfaces are shown atop the taped particle detector head entrance apertures for ions (bottom left) and electrons (bottom right).}
    \label{fig:ext_view}
\end{figure}

The EPD instrument consists of a pair of solid state, telescopically-collimated detector heads to detect electrons and ions within designated fields of view, shown in Fig. \ref{fig:epd_fovs}. As described in the introductory sections, the detector was developed to collect differential-directional particle flux spectrograms, designed for the target magnetospheric environments of Earth's outer radiation belt, ring current, and plasma sheet, over a relatively fast polar crossing time (6-7 minutes per science collection window). The instrument was also required to provide flux sensitivity and local particle pitch-angle resolution to observe the ionospheric bounce loss-cone and magnetically-mirroring populations across the high-latitude orbit track of the satellites (nominally $55^\circ$-$80^\circ$ magnetic latitude in the northern and southern hemispheres). At these latitudes, the expected ranges of energy-dependent particle count rates were estimated from previous mission observations and aggregated flux models, such as AE-9 and AP-9 \citep{Ginet2013}, informing the detector design choices. The two assembled EPD flight units (FM-A, FM-B) and a ground engineering model (EM3) are shown in Fig. \ref{fig:ext_view}.

The instrument generates energy-time spectrograms by counting individual particle impacts within a specified time window, commencing at the start of each satellite spin period. The energy of each particle is digitally binned into 16 discrete channels, covering energies $\lesssim$50 keV to $\gtrsim$5.8 MeV. The energy deposition efficiency of these channels in the underlying silicon stack is described in further detail later. The width of the channels $\Delta E$ are logarithmically spaced, being $\leq$50\% central energy $E$ of each bin (i.e., energy resolution of $\Delta E/E\leq0.5$). The time series of raw count histograms in each energy channel comprise the EPD Level-0 (L0) ``raw'' data. Upon post-processing on the ground, these data are transformed into Level-1 (L1) differential-directional number flux (``nflux'') and energy flux (``eflux'') spectrograms, as well as Level-2 (L2) pitch-angle spectrograms. Examples of EPD L0, L1, and L2 measurements in a typical ELFIN science collection, as well as descriptions of their configurable parameters, are presented later in this manuscript. Complete details of EPD mission operations, including data collection statistics and spatiotemporal coverage, can be found in a second companion paper \citep{Tsai2025}.

The ELFIN satellites were configured to spin with a nominal period of 2.8 s in a plane containing Earth's background magnetic field, allowing the EPD to trace a complete pitch-angle distribution once per spin period. Each spin period is subdivided into 16 contiguous integration intervals, referred to as ``spin sectors,'' with a nominal time resolution of 175 ms per sector (1/16 $\cdot$ 2.8 s). These sub-integrations represent the fundamental counting time of the instrument and were performed in parallel over identical time windows for both electron and ion observations. The real-time magnetic spin sector phase is determined using an on-board algorithm to monitor 80 Hz FGM data for zero-crossings in field components, described in the Electrical and Digital Design section. The corresponding time-dependent pitch-angles were determined on the ground by finding the best-fit spin phase to the particle data over its 6-7 minute collection interval. Alignment of the spin axis over time was maintained by periodic satellite reorientation (``re-or'') maneuvers using on-board magnetic torquer coils, during intervals of EPD non-operation. This procedure is described in the companion ELFIN mission publications.

The EPD instrument operates on the principle of solid state particle detection in doped silicon media. In this method, incident charged particles impact a silicon pn-junction, nominally producing one electron-hole per $\sim$3.6 eV of deposited energy \citep{Anderson1972}, for both negative and positive impact charge polarity. The energy deposited per unit length depends strongly on the impacting particle species and its initial energy, with atomic ions (e.g., protons) depositing substantially more energy in the same silicon track length than electrons, due to the much higher stopping power ($dE/dx$) of ions at the same kinetic energy \citep{pe_star}. The bulk volume of the silicon detector, referred to as the ``junction,'' is held under reverse-bias potential to form a depletion region, resulting in removal of mobile charge carriers in the background bulk. The resulting reverse-bias electric field also acts to transport charge carriers across the junction at rates higher than intrinsic recombination rates. In the absence of particle impacts, this biasing procedure leaves behind only a small temperature-dependent leakage (``dark'') current, whose magnitude and stability were characterized as part of flight detector selection.  

When incident electrons and ions impact the silicon detectors, they produce a track of liberated electron-hole pairs within the silicon, which drift across the junction at a rate dependent on their mobility and the applied reverse-bias electric field. These events manifest as a brief electrical current pulse (single nanoseconds to 100s of ns) emanating from the junction, corresponding to the energy-dependent particle impact charge transfer time. This signal is captured by a charge-sensitive preamplifier circuit with a characteristic integration time, chosen to be long enough to allow for sampling over all energies of interest but short enough to minimize noise. 

Because the impact charge pulse does not always have the same envelope (due to unequal electron-hole transit times, various impact track trajectories, charge diffusion, etc.), the preamp instead converts the total time-integrated input current (net charge) into a characteristic shaped pulse of known rise time (2.4 $\mu$s), with amplitude proportional to the deposited input charge (and thus, impact energy). The shaped output signal is sampled using an analog-to-digital converter (ADC), allowing the energy of the impact to be determined using pulse-height analysis (PHA). This is achieved by taking the difference $\Delta V$ between ADC readings at the start time and end time of the shaped output pulse. To first order, the deposited particle energy $E$ is related to this difference by $\Delta V = a_i E + b_i$, where $a_i$ and $b_i$ are the gains and offsets, respectively, of the $i^\textrm{th}$ particular detector electronics chain ($i=1,...,8$). The gains and offsets were determined experimentally using known radioisotopes as spectral calibration references, described later.

\section{Sensor and mechanical design}
\begin{figure}[!htb]
    \centering
    \includegraphics[width=\linewidth]{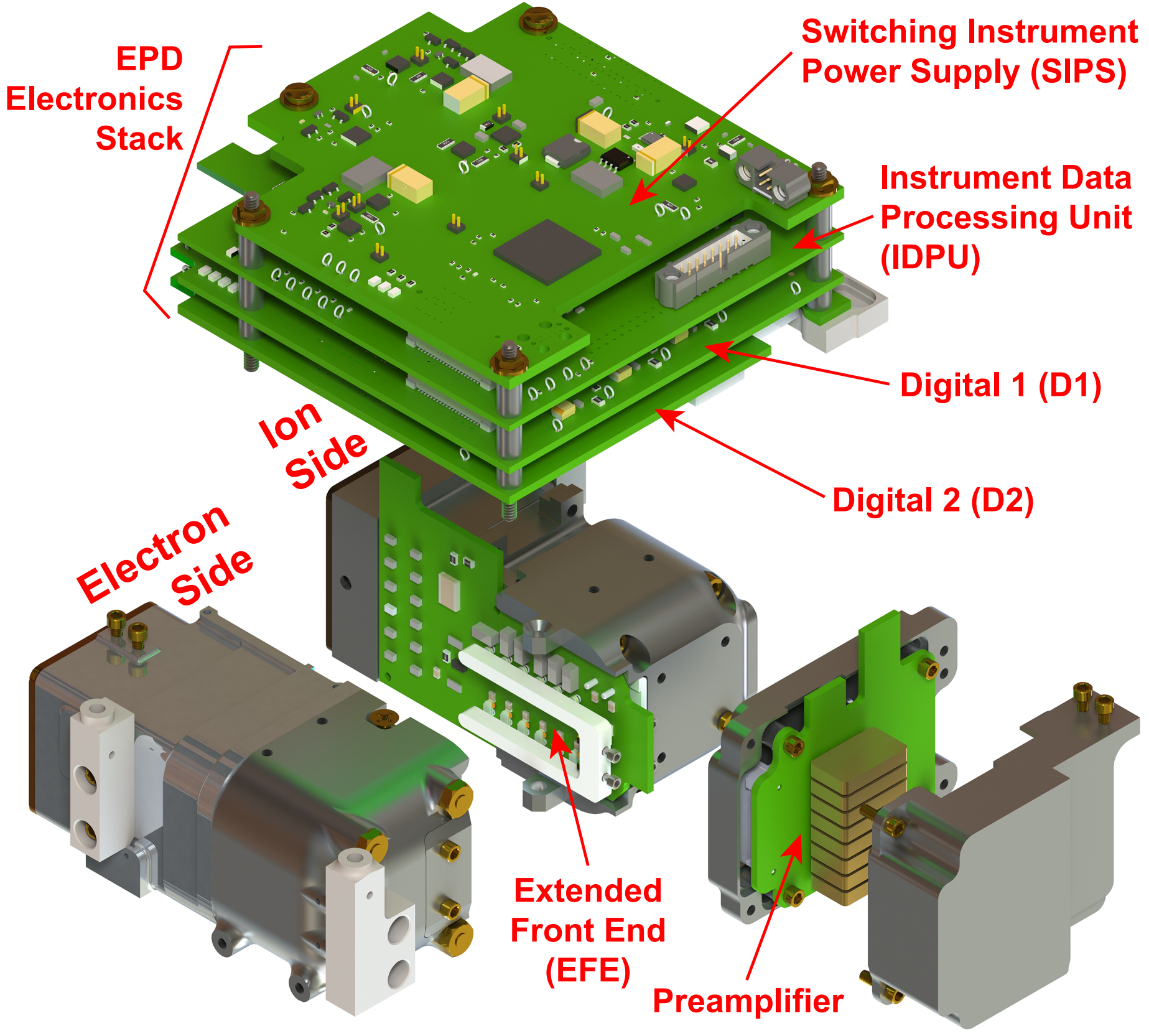}\\
    \caption{Exploded CAD view of EPDE/I detector head components. }
    \label{fig:epd_exploded}
\end{figure}

The EPD consists of two detector heads, which contain adjacent telescoping collimators for electrons (EPDE, or EPD-E) and ions (EPDI, or EPD-I), shown as an exploded view in Fig. \ref{fig:epd_exploded}. The detector heads share parallel optic axes with the CubeSat deployment direction, occupying the 1U panel opposite the deployment switches. Electron and ion sides capture the same look directions and pitch-angles, offset by 3.9 cm relative to their bore-sights. The EPD and payload circuit boards fit within a 10$\times$9.8$\times$8.4 cm$^3$ volume, weighing 939 g. 

\begin{figure}[!htb]
    \centering
    \includegraphics[width=\linewidth]{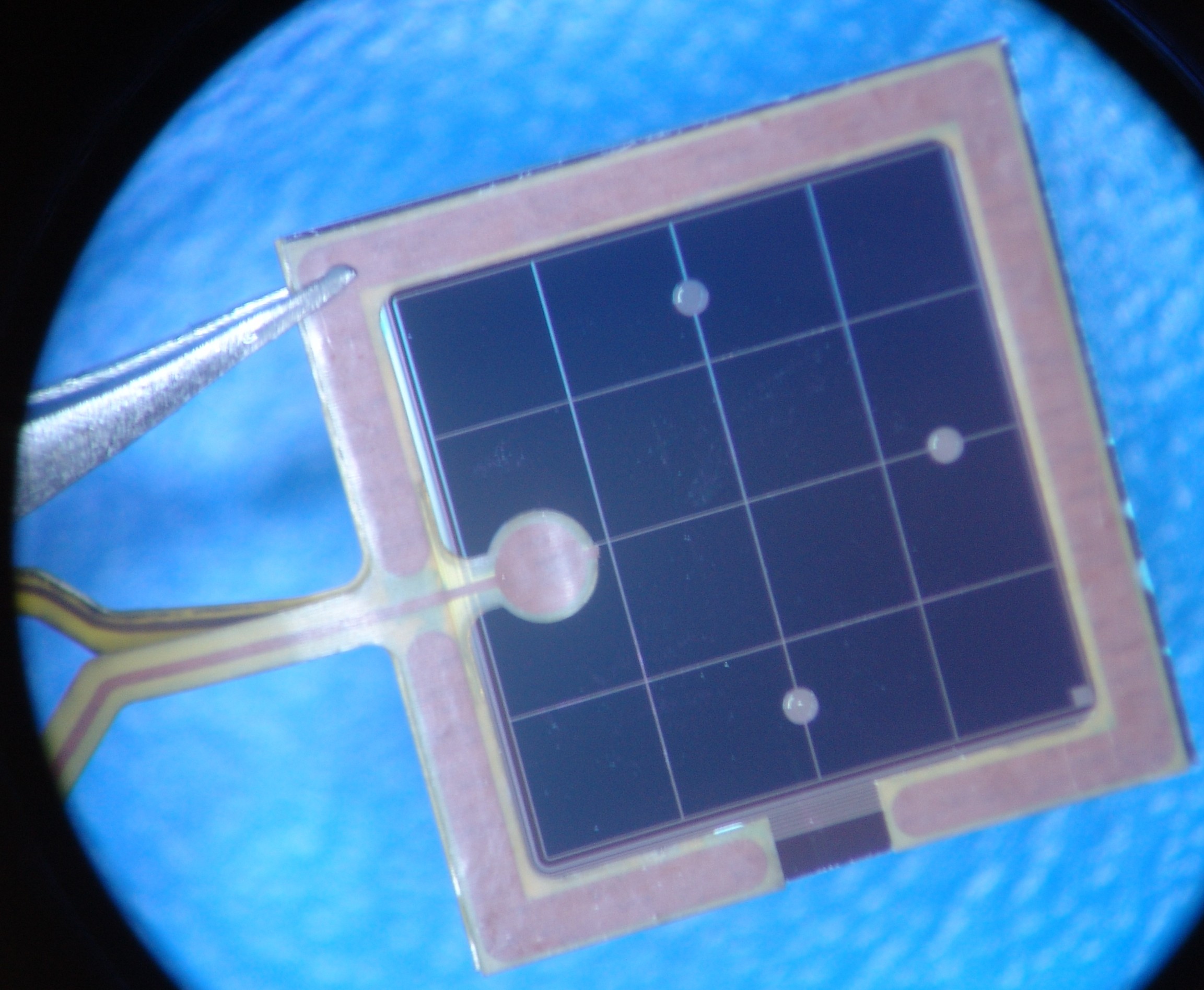}
    \caption{Micron MSX-03 silicon detector, showing junction side (minimal surface metalization) with 10 mm x 10 mm active area on a 13 mm x 13 mm chip. The design incorporates a charge-collection grid and conductive epoxy bonding points. The detector uses a kapton flexi adhered to perimeter with non-conductive epoxy. Typical wirebond contacts were avoided to minimize channel summing losses through the stack of silicon chips. Opposite face (not shown) is uniformly flashed with 1000 Å of aluminum for light rejection in the EPDI detector head.}
    \label{fig:det_single}
\end{figure}

A total of 8 silicon detectors (6 for electrons, 2 for ions) are used in the design. The detectors were of the MSX-03 series produced by Micron Semiconductor, with unit-specific production thicknesses, doping profiles, and reverse-bias depletion voltages determined by Micron prior to delivery. Each detector entrance contact contained a thin layer of polysilicon and bonded conductive grid to maintain bias electric field uniformity and charge collection, shown in Fig. \ref{fig:det_single}. The reverse side was flashed with $\sim$1000 Å aluminum to mitigate light contamination, with surface connection by a conductive epoxy bond point. The detectors possessed electrical contacts for the junction and ground interfaces, with kapton flexi pin-style leads used to mate with the electronics front-end interface. The detectors included a guard ring held at ground potential. The capacitance of each fully-depleted (reverse-biased) detector was modeled as a rectangular parallel-plate with silicon dielectric constant.

\subsection{EPDE detector head}
\begin{figure}[!htb]
    \centering
    \includegraphics[width=\linewidth]{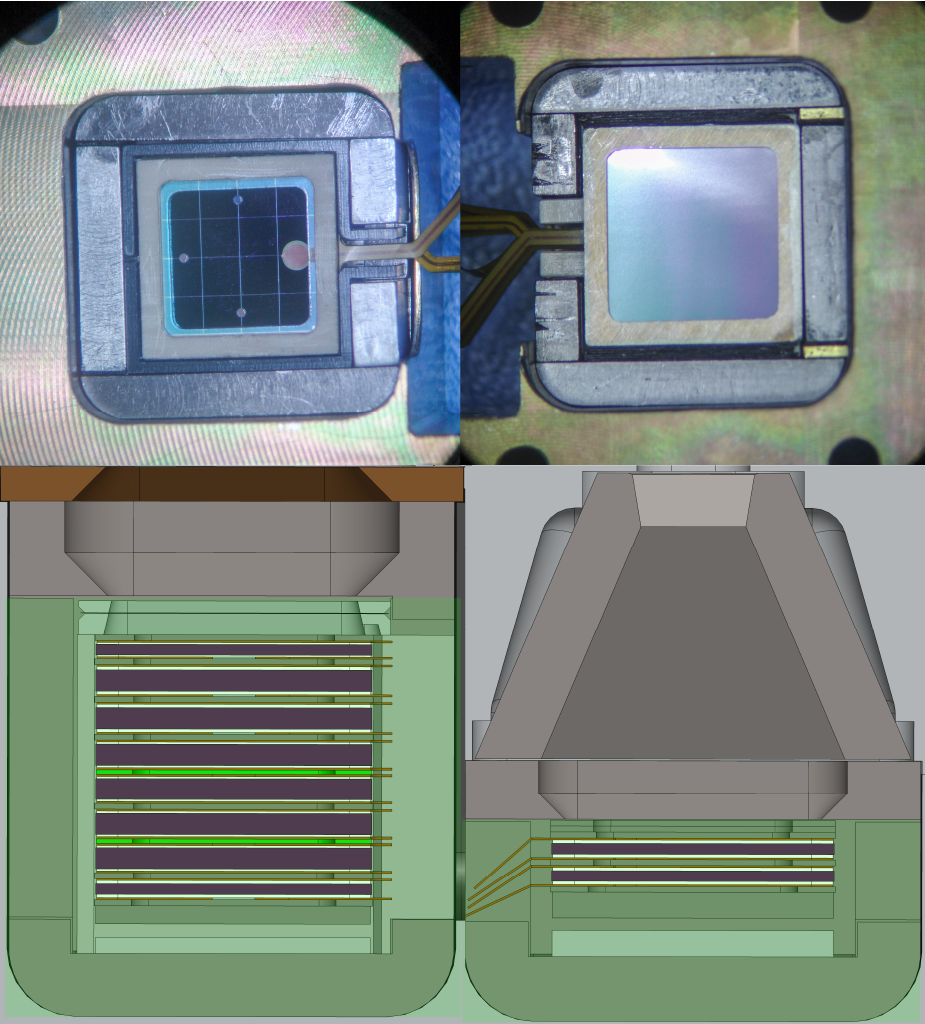}
    \caption{Top-down view of the EPDE (left) and EPDI (right) detector stacks with cross-sectional view at the bottom in green. The E1 and P1 (stack top) detectors are shown, where it is seen that the P1 detector has its flashed side facing outward to block sunlight. EPDE lexan foil is not shown.}
    \label{fig:detectors}
\end{figure}

The electron detector head consists of a stack of six silicon detectors enclosed in a bronze shielding vault with tantalum retainers, shown in Fig. \ref{fig:detectors} (left). The front face of the vault is exposed to allow electron flux transmission to the silicon detectors. The detectors possessed thicknesses along the optic axes of 500 $\mu$m (E1), 1000 $\mu$m (E2), 1000 $\mu$m (E3), 2000 $\mu$m (E4-ganged), 2000 $\mu$m (E5-ganged), and 500 $\mu$m (E6). Each detector in the stack shared a common active area of 10 mm x 10 mm on a 13 mm x 13 mm chip, with surface normal oriented along the optic axis, shown in Fig. \ref{fig:detectors}. The decision to use 6 detectors was to provide sufficient silicon stack thickness (0.7 cm) to sense electrons up to 5 MeV energies, while maintaining sufficiently low dark current and capacitance so as to resolve the low-energy portion of the target science range. Each detector is separated by $\sim$10 thou (254 $\mu$m) PEEK spacers, with a base-plate wave washer used to compress the stack in a rigid configuration. The washers were procured commercially, with quality control in the washer spring force and loading hysteresis proving to be challenging obstacles, requiring extensive unit testing and screening prior to component integration.

The detector flexi feet are exposed radially inward toward the optic axis through the vault to allow interfacing with the front-end electronics, providing bias voltages and capacitive coupling to the preamps. The detectors are nominally reverse-biased above full depletion at $\sim$80 V for E1 and E6, and $\sim$160 V for E2, E3, E4, and E5. The bias voltages are configurable through the instrument command interface (described later). In practice, the operating voltages were set above the full-depletion level reported by Micron, allowing margin for potential sag or drift in the detector bias supply.

The E4 and E5 detectors were formed using a special process to merge two 1000 $\mu$m detectors into a parallel electrical and mechanical configuration. This procedure is referred to as ``ganging,'' which allows two thinner silicon detectors to act as a single thicker silicon detector, effectively halving the required bias voltage at the cost of increased capacitance and dark current. This trade was deemed acceptable based on the requirement of $\Delta E/E < 0.5$, as the E4 and E5 detectors nominally measure $>$1 MeV electron energies. 

\begin{figure}[!htb]
    \centering
    \includegraphics[width=\linewidth]{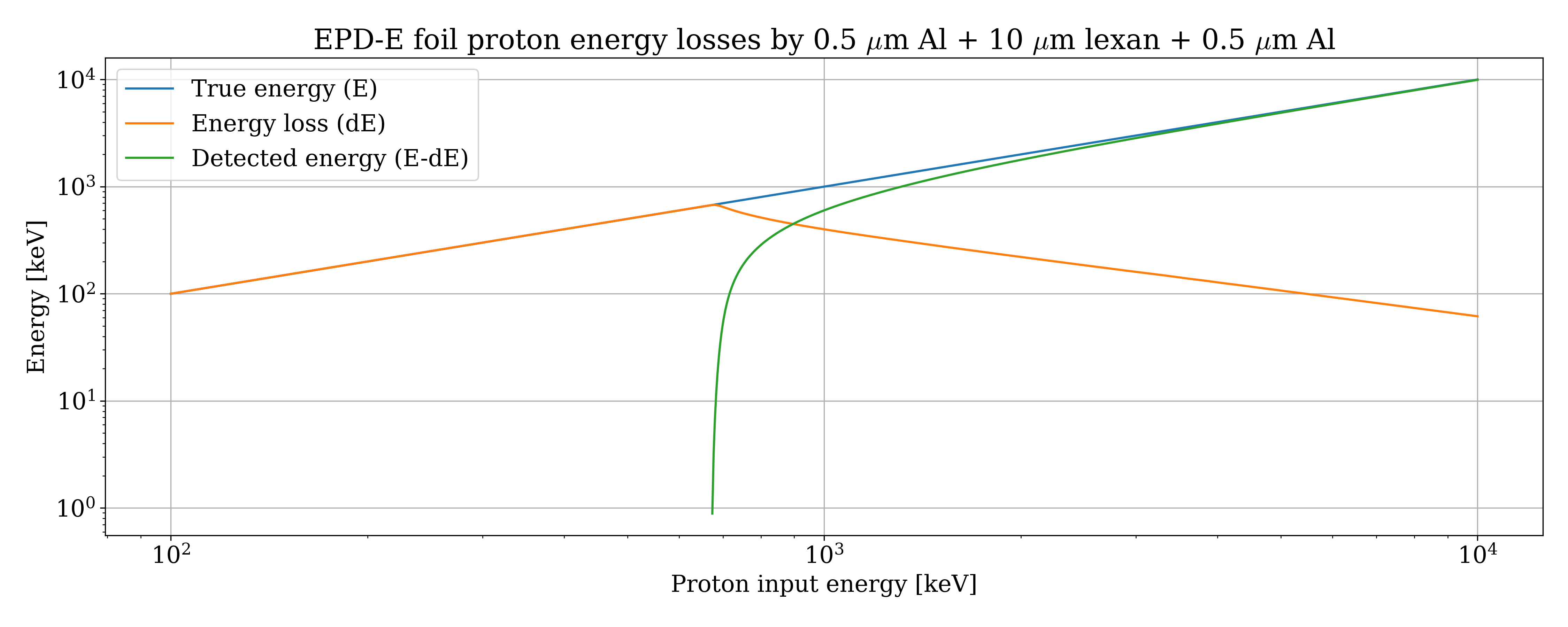}
    \caption{Proton penetration through the aperture foil into the EPDE silicon stack as a function of incident energy, computed using the Continuously Slowing-Down Approximation (CSDA). Protons are completely rejected below $\sim$700 keV, appearing at reduced energies above this threshold.}
    \label{fig:proton_srim}
\end{figure}

To mitigate contamination by sunlight and protons (below $\sim$700 keV), the EPDE included an aperture foil consisting of a 10 $\mu$m Lexan polycarbonate layer sandwiched between two layers of 0.5 $\mu$m aluminum, shown in Fig. \ref{fig:detectors}. Protons above $\sim$650 keV possessed sufficient energy to overcome the losses within the foil barrier, appearing at reduced energies to the EPDE detectors as shown in Fig. \ref{fig:proton_srim}. In order to register on the EPDE, the proton must have at least 50 keV remaining energy after the impact; thus, $\sim$700 keV is taken as the minimum penetration energy. Based on comparisons with EPDI data (described below), proton contamination in electron channels tended to be negligible except for extreme geomagnetic events, such as Solar Energetic Particle (SEP) conditions, or crossing the South Atlantic Anomaly (SAA), which can each be easily identified in the data as a separate isotropic energetic population above the main (typically lower-energy) electron phase space density. 

\begin{figure}[!htb]
    \centering
    \includegraphics[width=\linewidth]{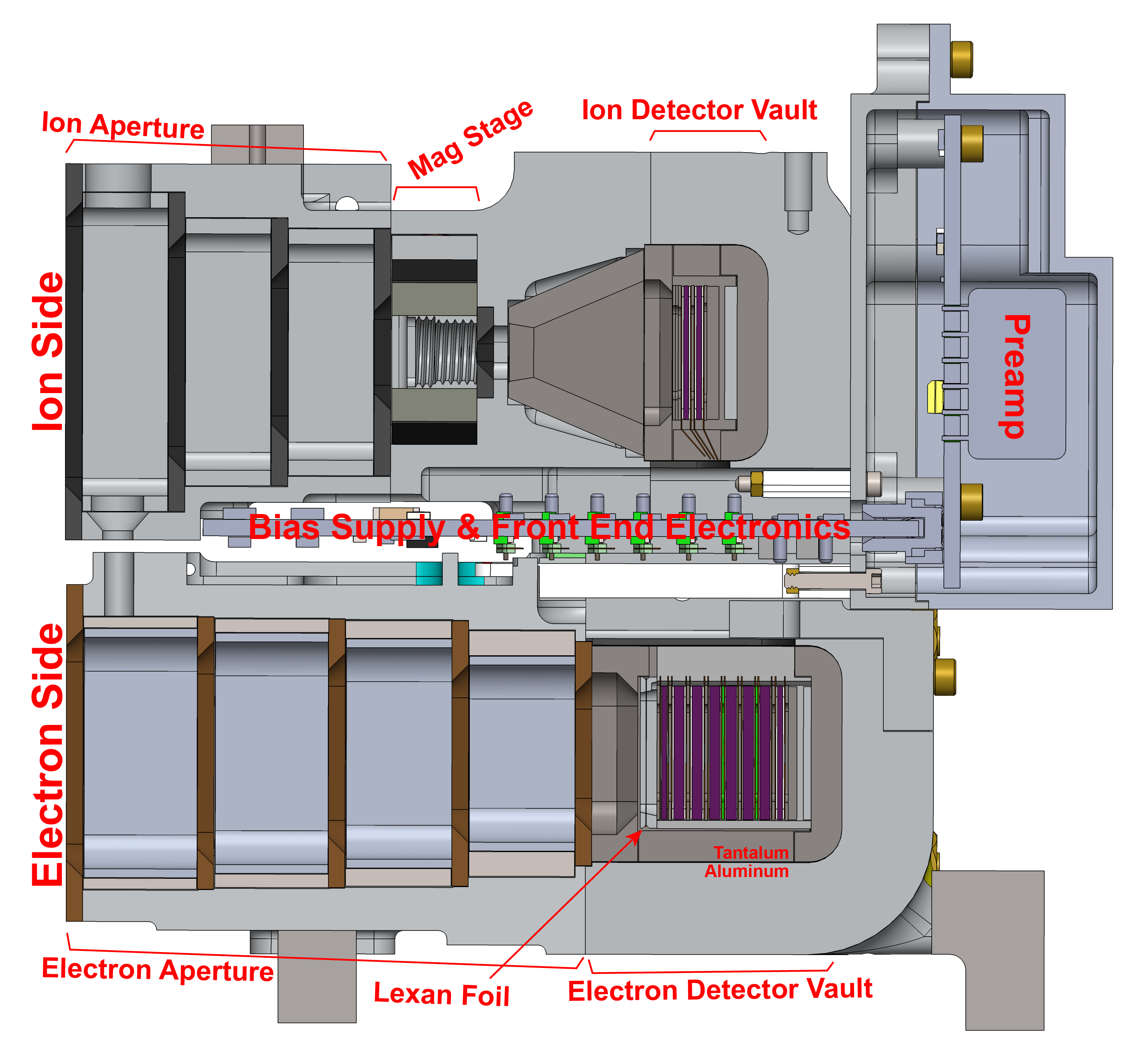}
    \caption{Annotated cross-sectional cut of the EPD detector heads. Particles enter at entrance apertures, collimated by the Field of View knife-edges and subsequently filtered by the mag stage (EPDI) or Lexan foil (EPDE). Each detector stack is biased and sampled by front-end electronics.}
    \label{fig:epd_cross}
\end{figure}

Incident particle fluxes are collimated using a telescoping $22.5^\circ$ field-of-view (FoV) with entrance aperture, shown in Fig. \ref{fig:epd_cross}. Collimation is achieved through pyramidal arrangement of knife edge scattering surfaces oriented perpendicular to the optic axes, composed of phosphor bronze, and blackened using Ebonol-C to reduce stray light scattering. Particles out of the intended field impact the knife edges, tending to either undergo specular reflection or deposit their energy within the surface. Particles in the intended view field pass through an aperture of area 4.36 cm$^2$, placed at a distance of 1.5 mm from the Lexan foil. This combination of focal distance and aperture area results in a geometric factor (angular-area flux acceptance) constrained by the collimator geometry rather than the detector area. 

This detector head design rejects the vast majority of out-of-field electrons and ions of magnetospheric origin. However, even with dense shielding, extremely energetic particles ($>$50 MeV) are capable of penetrating into the detectors from outside of the intended FoV. In such cases, the contamination is instead mitigated using detector stack coincidence logic (described in further detail in the electrical/digital section), implemented in the front-end digital electronics. During a particle impact event, this is achieved by checking whether a physically plausible combination of detectors in the stack received deposited energy. For example, a particle depositing energy on E1 and E2 but not E3-E6 is a physically possible trajectory within the intended FoV (arriving head-on). However, a particle depositing energy on a middle detector only (e.g., E2 alone) suggests a particle that penetrated the instrument's side shielding rather than entering head-on. Such counts are automatically rejected by the system, based on their coincidence pattern. In practice, only a small fraction of counts tend to be rejected this way, owing to the number fluxes of particles at these energies being orders of magnitude lower than those in the EPD's intended energy measurement range.

\subsection{EPDI detector head}
The EPDI is the corresponding detector head for atomic ions. The instrument did not possess the capability to resolve the composition of ion species but was intended primarily to detect protons, based on the polar LEO orbit profile of the ELFIN mission. The EPDI design is similar to that of the EPDE, but differs in several key respects. First, it makes use of only two silicon detectors: P1 (500 $\mu$m) and P2 (500 $\mu$m), biased at 80V. This is because protons deposit their energy much more effectively per unit length in silicon than electrons over the 50 keV to multi-MeV range, and thus require less stack thickness. The P1 detector is capable of sensing protons at energies up to at least 6 MeV, and accounts for the entirety of ion flux measurements. The P2 detector is used entirely for anti-coincidence logic, in order to reject contaminant electrons at energies exceeding 500 keV, which could not be pre-filtered by the entrance aperture.

\begin{figure}[!htb]
    \centering
    \includegraphics[width=\linewidth]{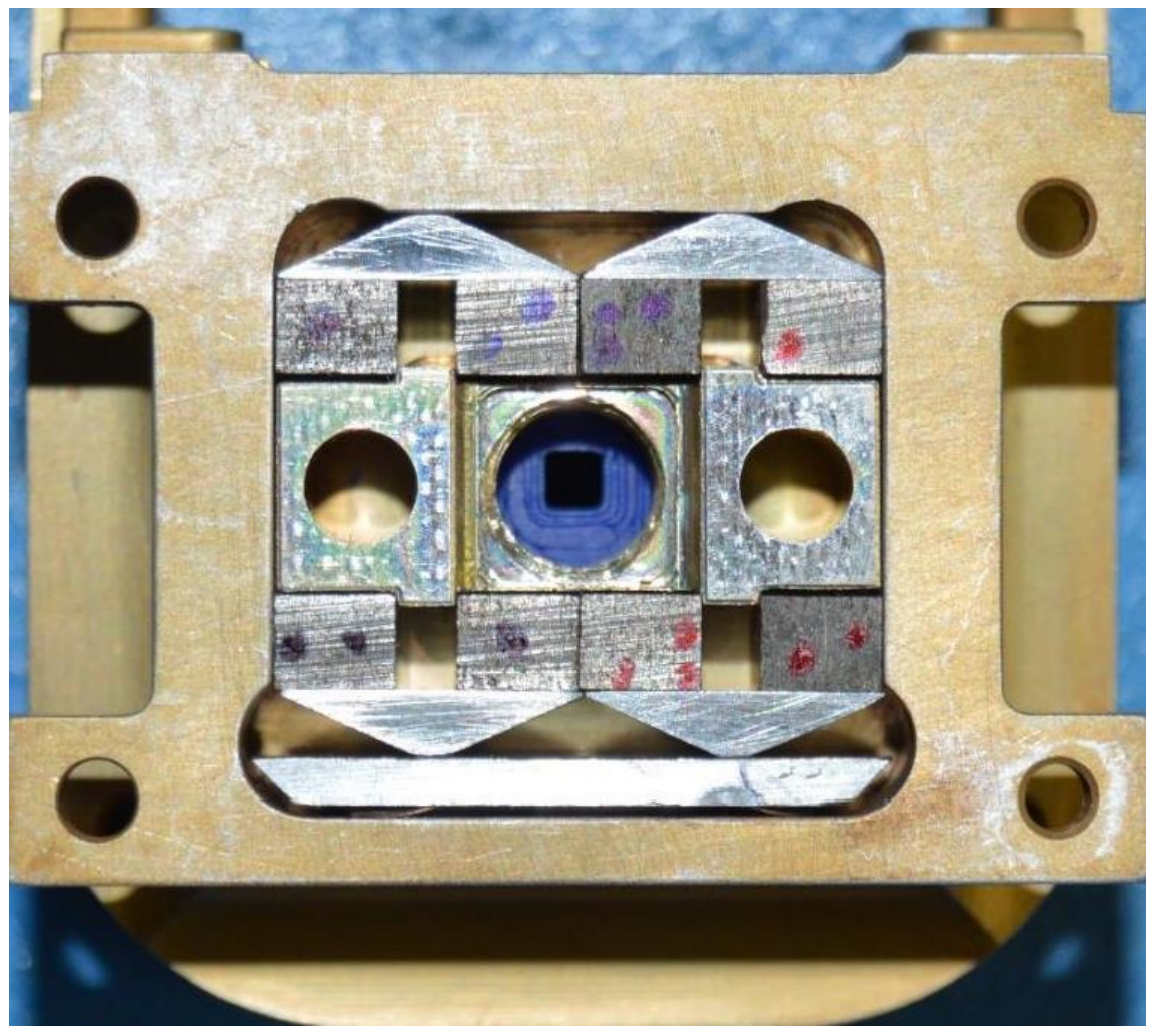}\\
    \includegraphics[width=\linewidth]{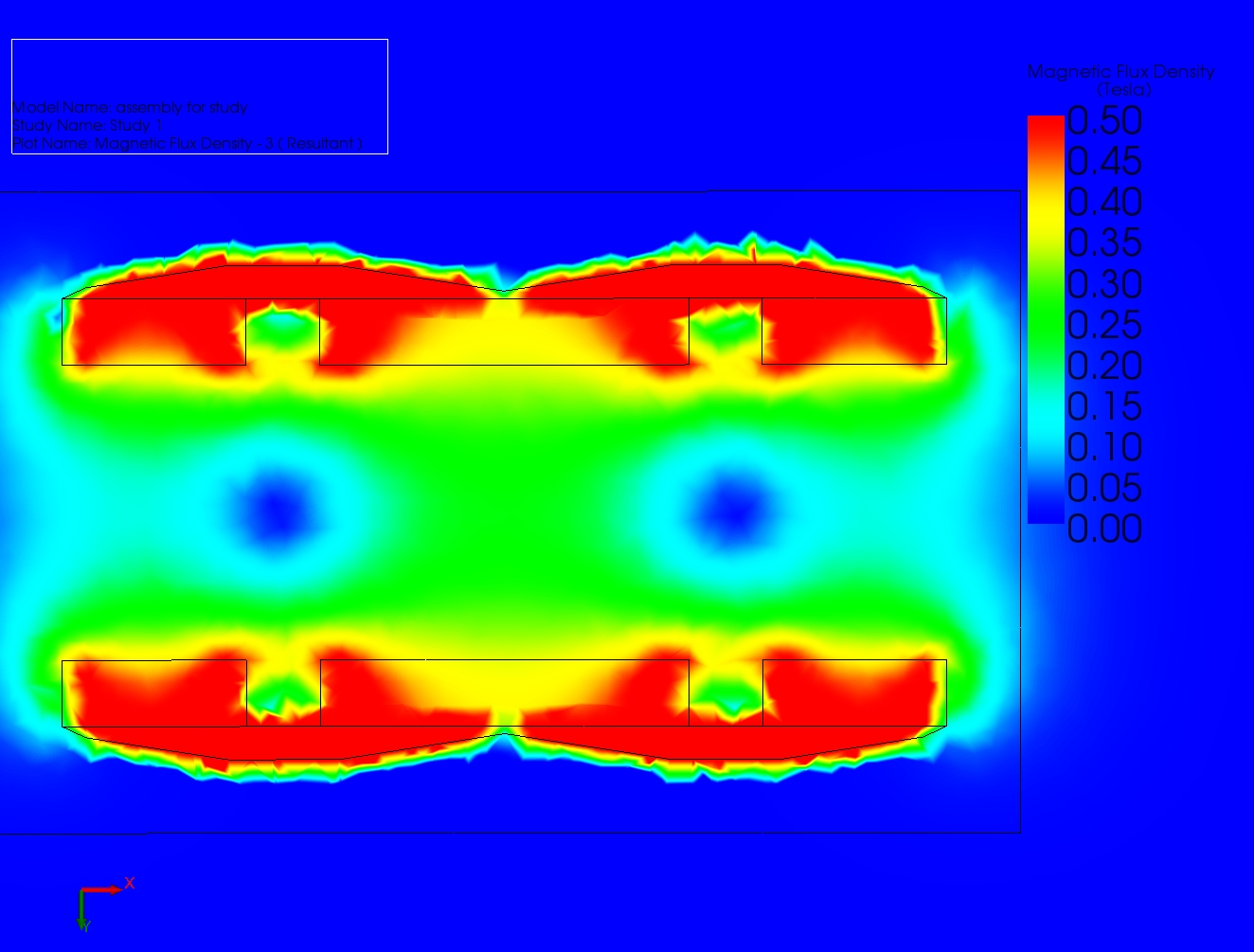}
    \caption{Cross-section of the magnetic broom stage (``mag stage'') aperture of the EPDI (top) including yokes and cores, and corresponding simulated magnetic field magnitude spatial profile (bottom). Electrons which pass through the central axis are lost below $\sim$500 keV energy.}
    \label{fig:mag_stage}
\end{figure}

The EPDI collimator similarly possessed an entrance aperture geometry with pyramidal telescoping knife edges to impart the desired field of view. Compared to EPDE, a significant EPDI mechanical difference is the replacement of the Lexan aperture foil with a magnetic broom filtering stage (``mag stage''), shown in Fig. \ref{fig:mag_stage}, designed to reject electrons from the field of view. This was achieved by enclosing the entrance aperture in region of locally strong magnetic field, designed to deflect a preferential energy range of charged particles via the Lorentz force. Because electrons possess a much smaller mass than protons, they are preferentially deflected into the side wall under the same magnetic conditions, typically being lost prior to depositing any energy in the P1 detector. The distance between P1 and the collimating focal point within the mag stage was 16.75 mm, with a correspondingly smaller circular entrance pinhole of radius 4.33 mm, which established the geometric factor of the instrument. The proportionally smaller aperture area compared to the EPDE corresponds to the prediction of higher flux levels of protons at 10s to 100s of keV energies in the typical orbit environment, with the goal of limiting count rate saturation and maximizing dynamic range.

The mag stage used in the EPDI achieved a maximum field strength of 0.47 T within the aperture (0.28 T average gap field), sufficient to deflect electrons up to $\sim$500 keV into the side wall. The deflection efficiency was verified using a combination of particle tracing simulations and radioactive sealed source testing (both described later). In order to prevent reflections of the deflected electrons, the mag stage walls were serrated to provide surfaces for repeated local scattering, and thus energy loss prior to reaching P1. The magnetic field was produced using trapezoidal-shaped magnetic yokes made of VACOFLUX 50 (a soft-magnet with high cobalt content) to guide and retain the magnetic flux generated by the N40-grade neodymium magnets in an octupolar configuration, which minimized contamination of the nearby Fluxgate Magnetometer and EPDE collimators. The mag stage was sized to sufficiently minimize electron contamination to below 7\% while optimizing thermal stability and field leakage, taking orbit-average fluxes from AE-8/AE-9 as representative \citep{Ginet2013}. Magnetic characterization testing ultimately showed an effective dipole moment of $8\pm2\text{ mA m}^2$ at a distance of 2 cm away.

\section{Electrical and digital design}
% The EPD instrument relied on several electronics boards, including for power, detector biasing, signal amplification, data acquisition, system control, and telemetry. The design specifications and methodology are described below. The boards are each interconnected through a stacked mezzanine interface, containing both analog signals and a digital bus, allowing inter-board communication via the Universal Asynchronous Receiver Transmitter (UART) protocol. A view of the boards in their entirety is shown in Fig. \ref{fig:epd_boards}.

\begin{figure}[!htb]
    \centering
    \includegraphics[width=\linewidth]{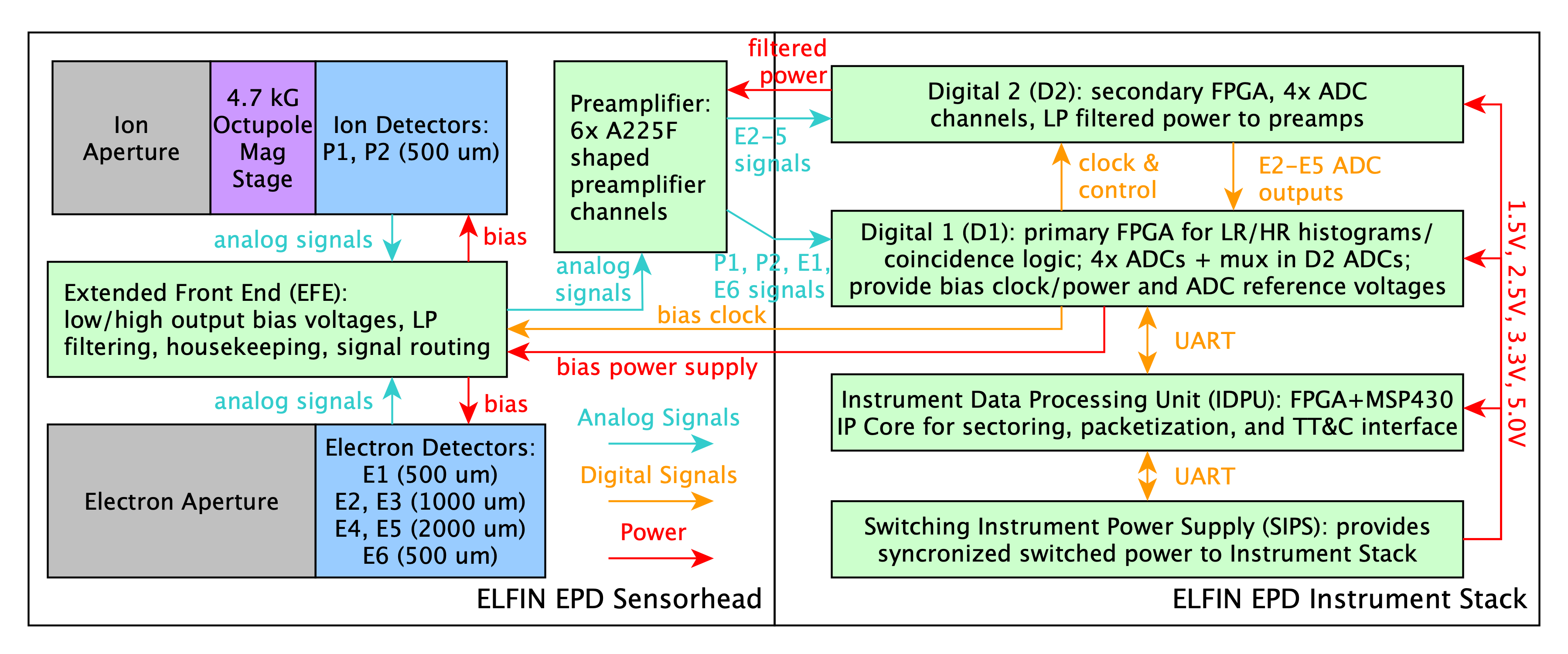}
    \caption{Block diagram of top-level EPD electrical interfaces. Power rails and bias lines are shown in red; analog detector signals shown in teal; and digital signals in orange.}
    \label{fig:electrical_top_level}
\end{figure}

\begin{figure}[!htb]
    \centering
    \includegraphics[width=\linewidth]{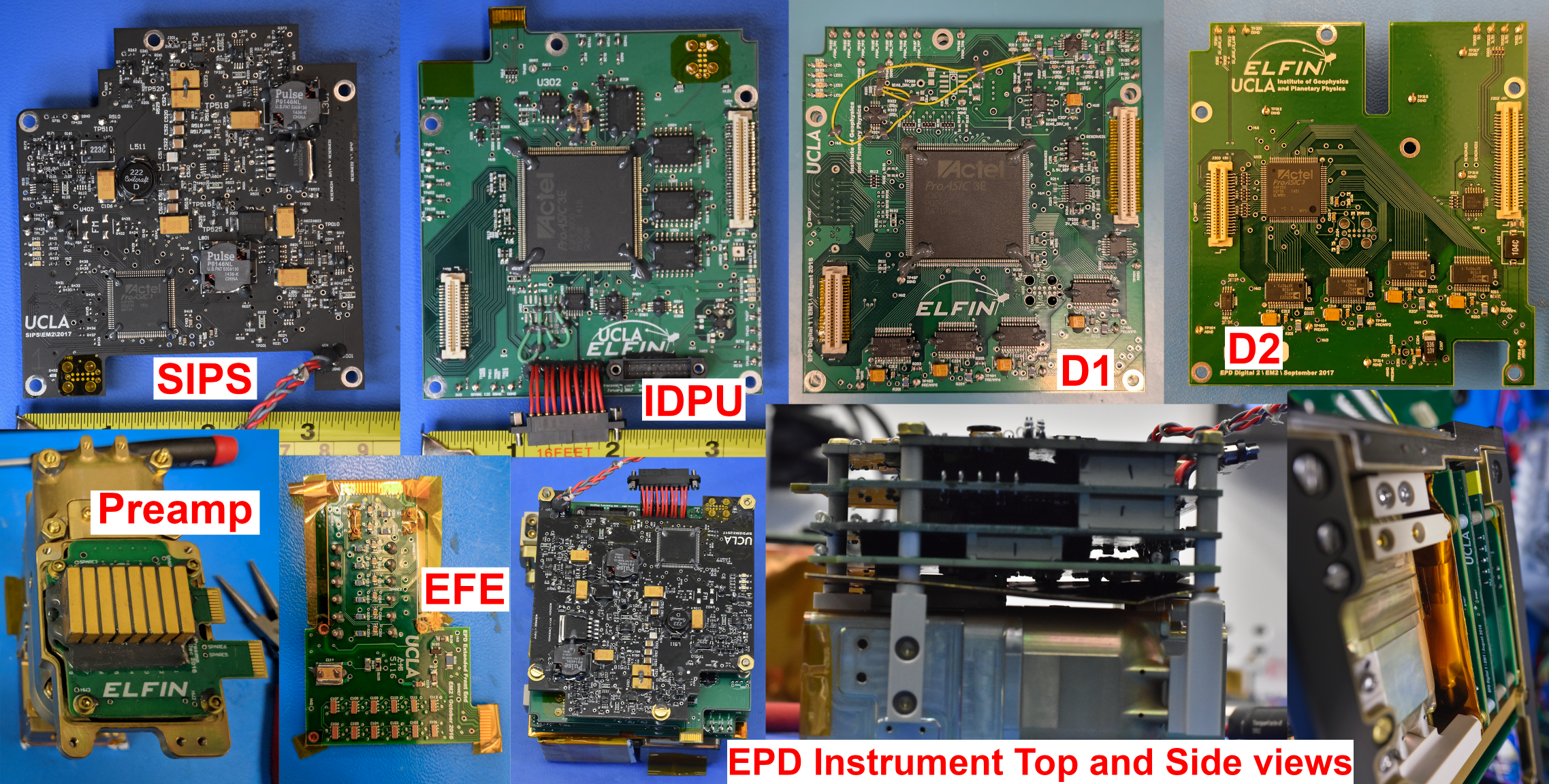}
    \caption{Montage of EPD electronics boards, including the Switching Instrument Power Supply (SIPS), Instrument Data Processing Unit (IDPU), Digital-1 (D1), Digital-2 (D2), Preamplifiers (Preamp), and Extended Front-End (EFE) boards. Assembled stack views are shown in the bottom right.}
    \label{fig:epd_boards}
\end{figure}

The EPD used a stack of four compact printed circuit boards (PCBs) sitting atop the detector heads. These boards provided power switching and housekeeping through the Switching Instrument Power Supply (SIPS); interface, control, sectoring, data processing, and telemetry functions through the Instrument Data Processing Unit (IDPU); and data acquisition and digitization through the Digital 1 and Digital 2 boards (D1 and D2, respectively). Within the sensor-head assembly, the extended front end (EFE) supplied detector bias voltages and provided front-end signal filtering. The EFE was connected to a preamplifier board mounted aft of the ion-detector vault. Multilayer insulation (MLI) blankets were installed above and below the detector heads to provide thermal insulation and maintain the detectors at low operating temperatures. The four-board stack was interconnected through stacked mezzanine connectors carrying both analog signals and a digital bus, with inter-board communication implemented using the Universal Asynchronous Receiver/Transmitter (UART) protocol. The top-level electrical signal block diagram is shown in Fig. \ref{fig:electrical_top_level}, while the complete board assembly is shown in Fig. \ref{fig:epd_boards}.

\subsection{EPD payload circuit boards}
\subsubsection{Extended Front-End board}
The ``Extended Front-End'' (EFE) provided the bias supply for the silicon detectors. This was achieved using a Villard Cascade voltage multiplier (capacitive network), where the input to the voltage multiplier was a bias clock signal at configurable frequency, using a MAX256 H-bridge driver. The optimal driving frequency was determined using spectral transform methods in the laboratory to minimize noise in the frequency band of particle impact pulses. The EFE also mated directly with the two-pin detector flexi leads (``junction'' and ``ohmic'') with passive RC-filtering. The non-ground electrical nodes were coupled to the pre-amplifier board using 2200 pF capacitors. 

\subsubsection{Preamp board}
Each detector was allocated one Amptek A225 preamp, for a total of 8 (6 on the EPDE, 2 on the EPDI). The preamps had a nominal gain of 240 mV/MeV, and provided pulse-shaping capability used to determine particle energies via pulse-height analysis (PHA). The shaping rise time of the A225 preamp was nominally 2.4 $\mu$s, with a longer decay time. This rise time constrained the fundamental time resolution between two consecutive particle impacts and established its count rate saturation level. This shaping time is a fixed property of the A225, and was selected based on achieving at least $10^5$ counts/sec per the ELFIN science requirements. 

\subsubsection{Digital boards}
The EPD is equipped with a pair of digital control and data acquisition boards known as Digital 1 (``D1'') and Digital 2 (``D2''). D1 acts as the primary digital instrument controller and data link using a ProASIC A3PE-1500 FPGA clocked at 16.384 MHz, which cycled through analog-to-digital (ADC) detector reads at the nominal 2.048 MHz (main clock distributed sequentially between 8 detectors). The D2 board contained a slave ProASIC A3PE-300 FPGA which relayed four of the ADC outputs to D1 via mux selection. The shaped pulse outputs from the detectors were sampled by these boards, each supplied with their own set of dedicated 12-bit ADCs, nominally spanning a range of 0 to 3.84 V. This sampling provided approximately 4 keV of energy resolution per bit, and a time resolution of $\sim$0.5 $\mu$s, which was less than the preamp shaping time, in order to subsample pulse-height outputs during both rise time and decay intervals. The ADCs were divided evenly across D1 and D2, containing four each. 

The D1 FPGA provides a register interface exposed to the IDPU (see below) for instrument control, status queries, and data readout. The FPGA double-buffers 16 energy channels worth of counts in 24-bit registers, incremented upon the end of every counting event. The FPGA further contains 16 spin sectors worth of energy channel registers, plus an additional 16 reserved for special high-fidelity operating modes. Upon request, these registers are latched for sequential UART readout and storage in IDPU flash memory, for future compression and downlink. In parallel, the 16 channel counters for each spin sector are reset to zero for the next integration. Such readouts are nominally triggered by the magnetic spin sectoring algorithm within the IDPU, and typically occur once every $\sim$175 ms. 

\subsubsection{Instrument Data Processing Unit board}
The D1 FPGA was configured using a telecommand protocol, with commands and telemetry transmitted via a UART link to the payload Instrument Data Processing Unit (IDPU) at a rate of 115200 bits/sec. The IDPU contained the primary payload CPU, in the form of a TI MSP-430 SoftCore processor running atop a ProASIC A3PE-3000 FPGA. The IDPU board connected its FPGA through a mezzanine-style parallel electrical bus with each of the payload boards. The MSP was programmed using a custom payload operating system, with four total boot image slots provided in on-board SRAM, configurable on-the-fly via uplink. The MSP had two types of operational boot programs: (1) data acquisition, comprising the nominal real-time scientific data collection mode, which stored data in on-board flash memory for later retrieval; and (2) data compression and transmission, wherein previously collected EPD data was compressed, packaged, and transferred to the ELFIN Flight Computer (FC) for downlink. Each system boot was scheduled through the FC interface via ground operations and on-board payload scripts. 

\subsubsection{Switching Instrument Power Supply board}
The analog and digital electronics of the EPD were powered by a dedicated Switching Instrument Power Supply (SIPS) power board, providing 1.2V, 2.5V, 3.3V, 5V, and 8V references. The SIPS board was operated by an on-board ProASIC A3PE-1500 FPGA, providing configurability for bias voltage setpoints, driver frequency, voltage/current housekeeping, and fault protection for over-current and over-voltage events. The SIPS FPGA was commanded by the IDPU via the UART interface. 

\subsection{Particle impact event handling}
Particle impact events detected by the instrument, referred to herein as counting events (simply, ``counts''), are triggered based on the features of the shaped output pulse of each preamplifier, as well as their coincidence signature within the silicon detector stack. The D1 FPGA handles registration and processing of counting events, based on continuous output monitoring of each ADC channel. To register as the start of an event, the time-derivative of the ADC output is required to exceed a specified threshold (start), followed by a drop below a specified threshold (end).

Compared to previous missions, ELFIN EPDs utilize the novel approach of triggering impact events in the time-differential domain, rather than crossing an absolute signal threshold level. An advantage to this approach is that any prior baseline offset in the preamp output, such as from a recent particle impact occurring over a shorter timescale than the shaping time window, has minimized effect on the triggering of subsequent particle events. This method is useful in high count rate scenarios, though due to the time-differentiation, requires minimization of high-frequency noise. This was achieved using Fourier analysis, described below, to design filters around intrinsic noise sources within the electronics. Each silicon detector channel was configured with its own thresholds, allowing for individual tuning. The optimal event start and end thresholds were determined both on the ground, and as part of a noise thresholding campaign during on-orbit instrument commissioning. The values were configured as part of the on-board payload start-up scripts. 

Upon commencement of a particle impact event, the D1 FPGA measures deposited kinetic energy by taking the difference in the shaped preamp voltage pulse at start and end times corresponding to crossing rising and falling differential thresholds (determined in early on-orbit commissioning, described in Calibration). The system gains and offsets between measured pulse-height (ADC units) and incident particle energies were characterized statistically as part of the instrument ground calibration campaign for each detector, using radioactive isotopes prior to launch (described in the following section).

For particle impacts with sufficient energy to deposit across two or more stacked detectors, the system employs coincidence logic (EPDE) to determine the total deposited energy and whether the particle originated in the intended FoV; and separately, anti-coincidence logic (EPDI) to reject a suspected electron count. The counting event logic for EPDE and EPDI is decoupled, preventing one sensor from affecting the other. A counting event begins whenever any detector in the stack produces a pulse exceeding its configured starting threshold, and holds the detected peak signal amplitude until the counting event terminates. During the initial rise time of the pulse shaper, the FPGA continues to monitor other detectors in the stack for threshold exceedance. If another detector in the stack, such as from a more energetic particle impact, occurs during the original rise time, the count event window will extend up to an additional 2.4 $\mu$s. The overall counting event ends only when all active detectors have had their individual pulse heights captured. 

The above coincidence logic design allows more energetic (lower flux) particles to be preferentially counted in scenarios where low-energy (higher flux) particles may otherwise prevent their detection. However, this leads to paralyzable detector behavior \citep{Usman2018}, requiring a special energy-dependent count rate correction to be applied to the data at moderate to high count rates, described in the Data Products section.

\subsection{Magnetic spin sectoring}
EPD data are collected on a per-spin basis, subdivided into 16 individual spin sectors of equal collection time, based on the most recently determined period. The instantaneous spin period and zero phase are determined in real time by an algorithm within the IDPU MSP operating system, which monitors FGM three-axis magnetic field components at 80 Hz for characteristic rotational signatures. These include zero-crossings of a selected field component or its time-derivative, which mark the start time and phase of the first spin sector in each period. 

In flight, the condition used to mark the start of a new spin period was the filtered zero-crossing of the time-derivative. This allows any spin-modulated field component to be used even when the component itself does not cross zero. To account for the sensitivity of time-derivatives to noise, a configurable N-point finite impulse response (FIR) filter is used to isolate frequencies in the vicinity of the expected spin period (2-4 sec). A configurable minimum time between consecutive spin periods is also imposed to ensure that noise cannot prematurely trigger a new spin period in the middle of the true interval. On the ground, best-fit solutions to the sector phase offset (``phase delay'') from true zero (field-aligned) were determined as part of Level 2 data products.

\begin{figure}[!htb]
    \centering
    \includegraphics[width=\linewidth]{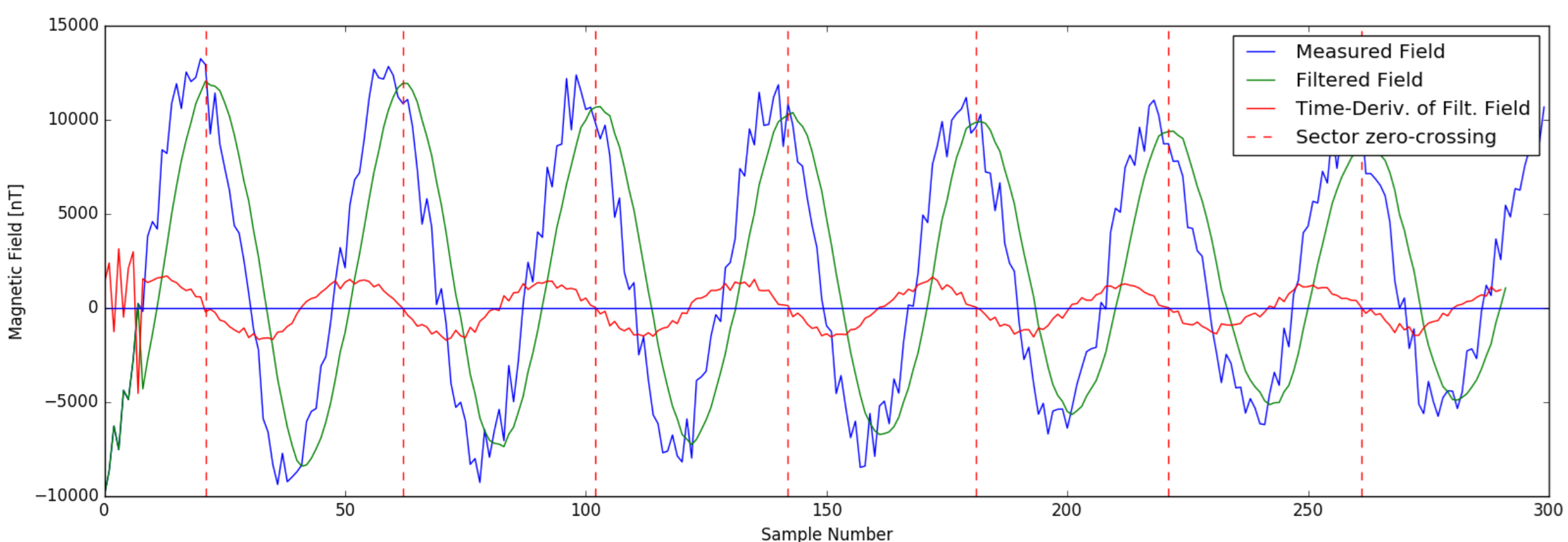}\\
    \includegraphics[width=0.5\linewidth]{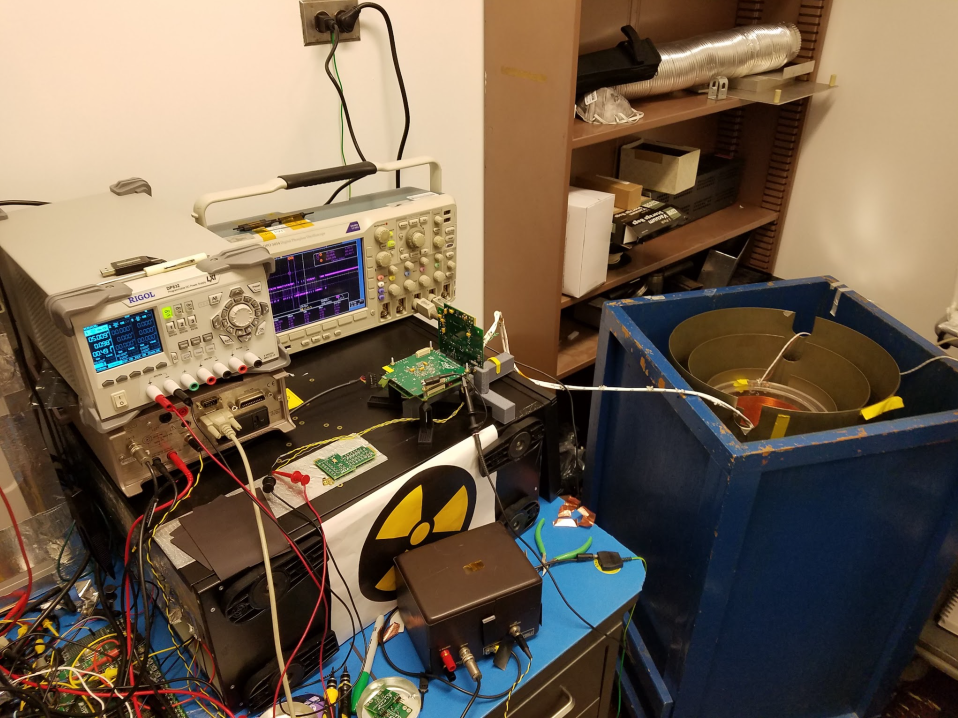}\includegraphics[width=0.5\linewidth]{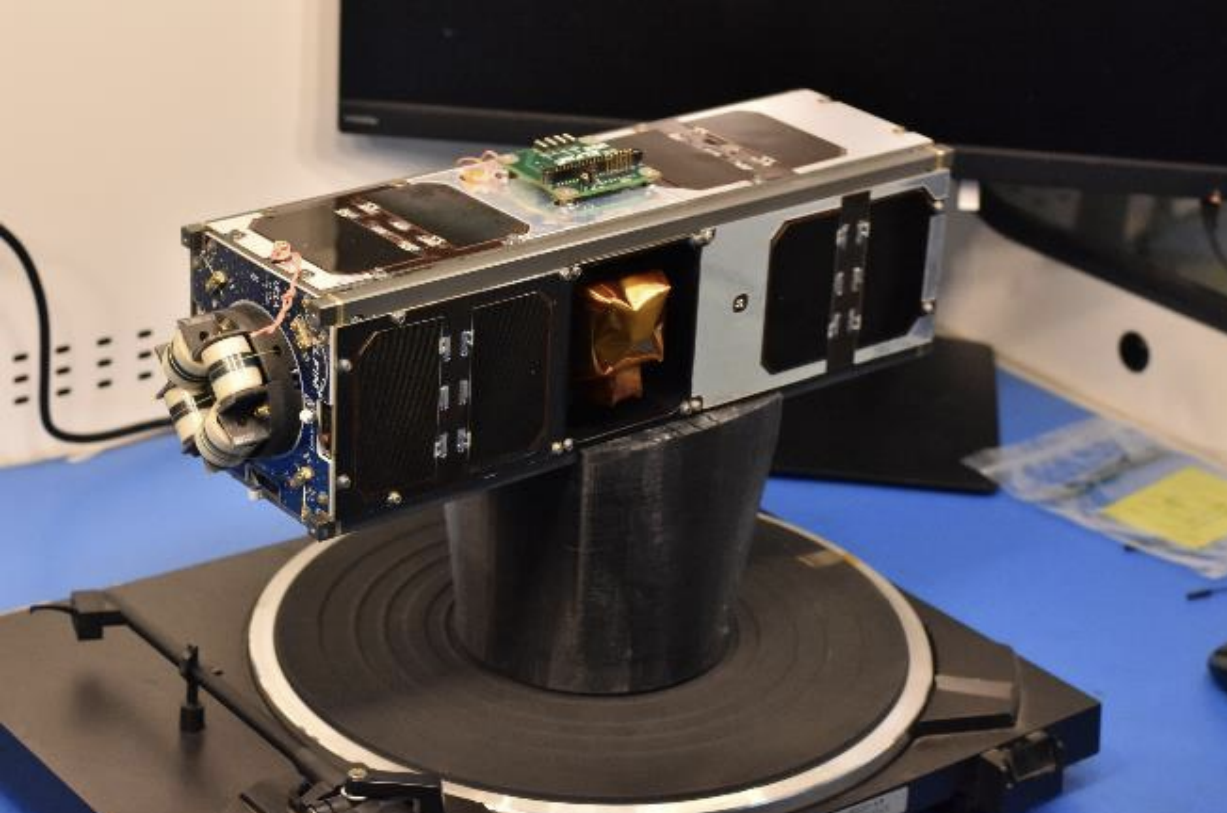}  
    \caption{Magnetic spin sectoring simulation (top), showing an example of filtered noisy FGM data and corresponding spin period triggering inferred by the algorithm. Several testing setups were used to verify the algorithm prior to flight (bottom), including direct injection of sinusoidal magnetic field components onto the FGM in a shield can (left), and rotation of the EM satellite relative to Earth's background field in lab (right).}
    \label{fig:spin_sec}
\end{figure}

A simulated example of the algorithm is shown in Fig. \ref{fig:spin_sec}. In this scenario, a sinusoidal magnetic field component at the nominal ELFIN spin period is processed using an identical method to that used on-board. A large amplitude noise signal is superposed on the sinusoid to demonstrate the robustness of the method. The signal is FIR-filtered, resulting in a highly-repeatable spin period, beginning at the vertical bars. A consistent phase offset is observed between the true and apparent zero-crossings, which is in practice extracted from the particle data by locating peaks in the particle flux density (corresponding to trapped particles with $\sim$$90^\circ$ pitch-angles) and finding the best-fit phase angle. 

\subsection{Data recording, compression, and transmission}
\subsubsection{Instrument startup and collection loop}
Real-time flight data collected during EPD observations is first configured and subsequently captured by the IDPU through the FPGA register interface to the D1 and SIPS boards. At instrument turn-on, the MSP operating system executes pre-uploaded flight macros to configure the instrument settings. These include enabling SIPS power supply voltage rails, setting individual detector thresholds, the number of spin sectors per period, detector bias supply clock/voltage setpoints, and the magnetic spin sectoring configuration. The bias supply is ramped in discrete steps from zero to its operating voltages to prevent excessive in-rush current and inductive effects. 

Upon execution of instrument startup, the MSP operates in an indefinite main loop carrying out the spin sectoring algorithm, sampling FGM 80 Hz data as it arrives. Upon detecting the first complete spin period, IDPU commands D1 to begin 16 consecutive integrations for the duration corresponding to the latest spin period. The time measurement is performed on D1 using a timer module to count clock cycles in the FPGA. Upon sensing the start of the next spin period, the previous spin period's particle data is read out through the register interface, and stored uncompressed in IDPU flash memory for later processing and downlink. This collection cycle repeats indefinitely until instrument shutdown. A real-time clock/calendar (RTCC) module on the IDPU tracks the timestamp of each individual collection, refreshed at startup by the flight computer.

The 16th spin sector is treated specially by being allowed a variable integration time. This is to account for the fact that variations in the spin rate are possible between each consecutive period, and the next collection interval may subsequently occur earlier or later than the preceding spin sectors. In practice, this resulted in a $\sim$10\% overestimation of the counts in this sector, which is accounted for as a fixed constant ``overaccumulation factor'' in the instrument calibration.

\subsubsection{Data compression and transmission}
To downlink data, the MSP is booted into a special compression and transmission mode. To minimize data volume, collected EPD data are subject to two types of lossy compression: count decimation and spin sector averaging. Count decimation served to reduce the bit depth of the channel counters from 24 to 8 bits. This is achieved using a lookup table mapping (LUT) from the original high-bit depth space to the 8-bit space. Separation between consecutive elements in the table was chosen to preserve the relative precision of the values. For example, low-count values from 0-10 are preserved as-is, while counts over 10,000 receive the highest decimation (losing precision in their least significant digits). Counts greater than 37,789 are placed in an integral bin in the 255th table index. Upon receipt by the flight computer, further lossless compression was performed prior to downlink.

\begin{figure}[!htb]
    \centering
    \includegraphics[width=\linewidth]{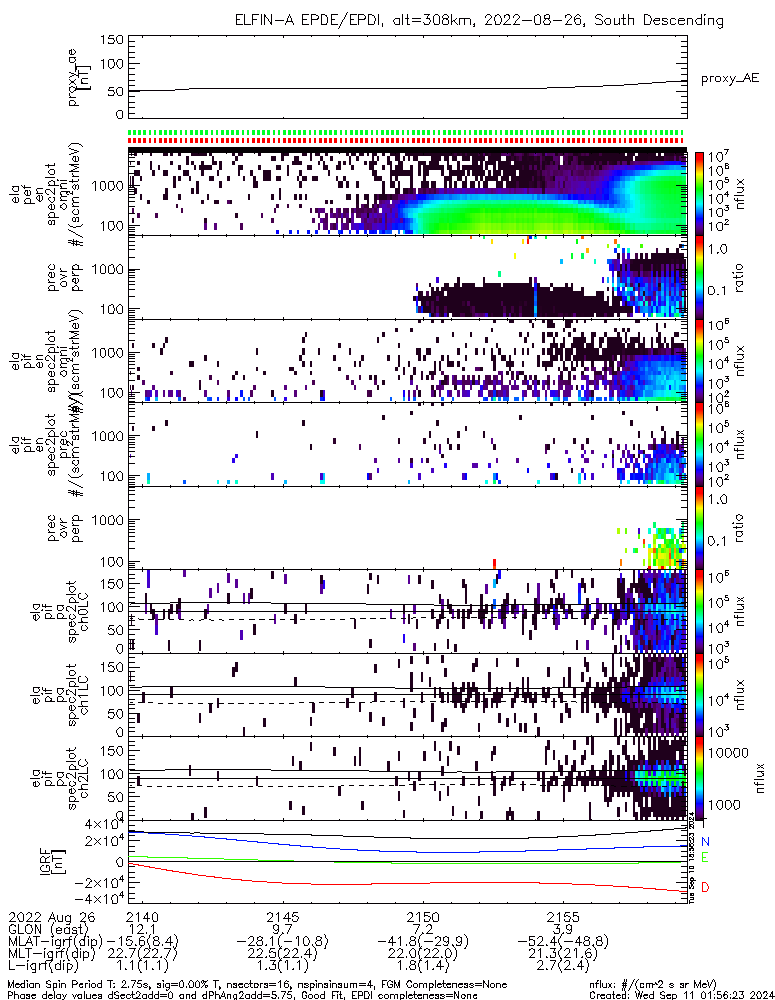}
    \caption{Example ELFIN Inner Belt Observations (IBO) collection. In this mode, the collection interval starts at much lower (equatorial) latitudes, collecting for over 20 minutes throughout the inner and outer radiation belts ($L>1.1$). In this mode, corresponding spin sector data in 4 consecutive spin periods are averaged to create a single pitch-angle bin. Panels are the same as in Fig. \ref{fig:overview_plots} (left).}
    \label{fig:ibo}
\end{figure}

Depending on the scientific objective, individual spin sectors may also be averaged on-board across multiple consecutive spin periods. This was performed primarily for the special Inner Belt Observations campaign (``IBO''), in which data collection occurred primarily at lower latitudes and over a much longer duration (spending more time at the same L-shells, allowing for time-averaging). In IBO collections, four consecutive spin periods were averaged on a per-sector basis. An example of an IBO science collection is shown in Fig. \ref{fig:ibo}.

\section{Calibration and performance characterization}
To characterize and select key components in the instrument design, several development and test campaigns were carried out. These included silicon detector evaluation, particle tracing simulations, radioisotope calibration, and noise thresholding analysis, described below. %Shock, vibe, and thermal vacuum (TVAC) campaigns were also performed prior to delivery.

\subsection{Detector selection}
To select the particular silicon detectors from the larger delivery batch to be used in the flight model (FM) EPDs, several criteria were evaluated: production quality, dark current performance, and long-term stability. Production quality was assessed by visual inspection under a microscope, as well as rigidity during mechanical handling. Of particular concern were the kapton flexi leads and periphery binding, which delaminated or broke in several instances. Where possible, the junction and ohmic contacts were also inspected for debris and integrity following vibe, shock, and thermal campaigns. 

\begin{figure}[!htb]
    \centering
    \includegraphics[width=\linewidth]{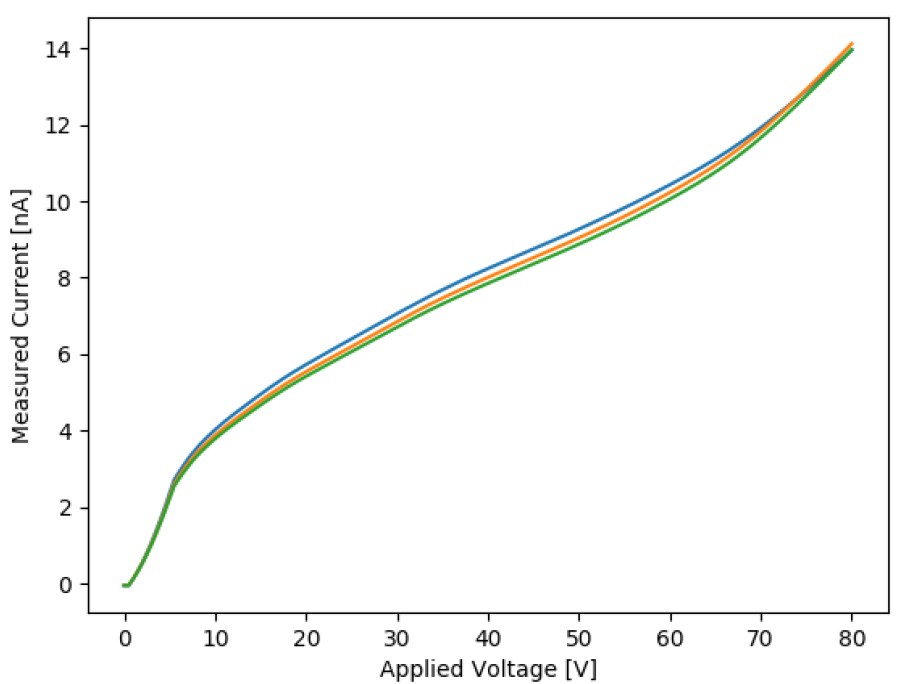}%{figs/test_2025-04-02_15_30_23.png}
    \caption{Example dark (leakage) current curves taken at room temperature for an MSX-03 detector delivered by Micron.}
    \label{fig:dark_current}
\end{figure}

The dark current performance of each detector was evaluated using a picoammeter. Each test run consisted of several IV-curve sweeps of detector bias voltage from 0 to a maximum value of 100V for 500 $\mu$m detectors and 200 V for 1000 $\mu$m and 2000 $\mu$m ganged detectors (all above the intended operating voltages). An example run on a 500 $\mu$m detector is shown in Fig. \ref{fig:dark_current}. The dark current for the flight detectors ranged from 5-10 nA for 500 $\mu$m units, 10-20 nA for 1000 $\mu$m units, and 20-40 nA for 2000 $\mu$m units. The effect of this current is primarily to broaden spectral peaks due to shot noise, which remained well below the instrument resolution requirements over the A225 preamp integration/shaping window.

The FM detector selection criteria were as follows: the lowest dark current 500 $\mu$m units were used as E1 and P1 (front), followed by P2 and E6 (back). The lowest dark current 1000 $\mu$m units were used as E2, then E3. Similarly for E4, then E5 with the 2000 $\mu$m (ganged) units. The remaining functional detectors were used in the engineering model (EM), used as a permanent flat-sat test bed during the operational lifetime of the ELFIN mission. 

We note that prior to launch, the E5 detector on FM-2 (ELFIN-B) failed post-integration. The particular cause is unknown, but resulted in a highly irregular, large-amplitude voltage waveform that did not correspond to any frequency in the system. The output of this detector was disabled for the entire mission, which according to simulation only affected the highest energy electron measurements ($>$3 MeV). Given the already low sensitivity of the instrument at these energies (see channel efficiencies below), the data were not noticeably affected compared to the ELFIN-A EPD.

\subsection{Energy and flux calibration: gains, offsets, and channel efficiencies}
To calibrate the gains, offsets, and energy channel flux sensitivity of the EPD detector chain, a combination of particle tracing simulations and laboratory radioisotope testing was employed. For simulations, Geant4 \citep{Agostinelli2003} was used in each case to inject particles both in and out of the EPD fields of view over a wide range of energies, depending on whether the objective is to calibrate energy or channel efficiency.

\begin{figure}[!htb]
    \centering
    \includegraphics[width=\linewidth]{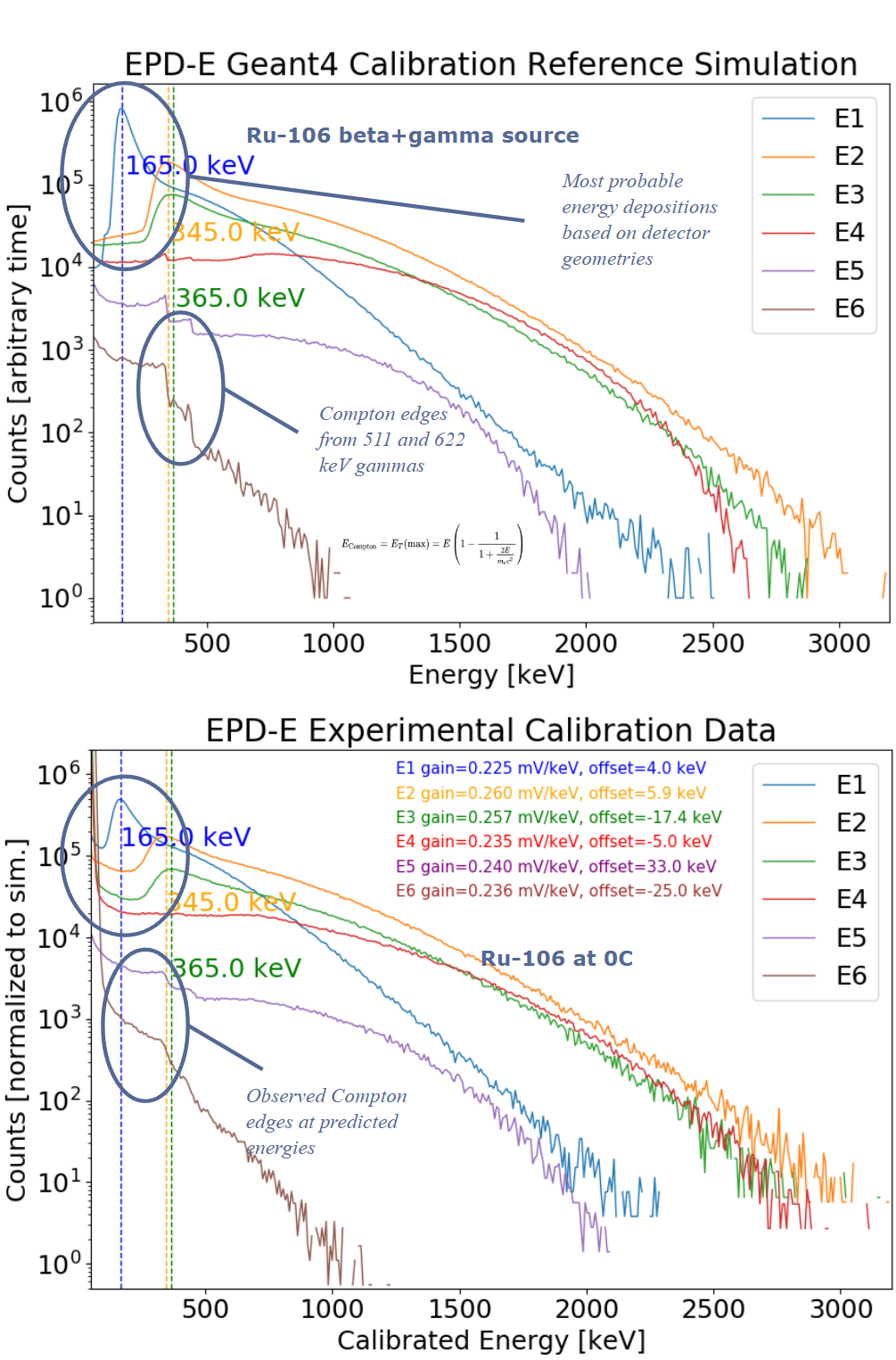}
    \caption{Example of detector energy calibration using the Ru-106 sealed-source isotope and FM-1 EPDE at 0$^\circ$C temperature. Top panel shows the detected energy spectrum predicted in each of the EPDE stack detectors (E1-E6) using Geant-4, while the bottom panel shows the results of the calibration campaign using an actual Ru-106 sealed source. Such curves were used to determine the gains (in mV/keV) and offsets (keV) of each detector channel assumed in flight by matching features such as the most probable deposited energies and Compton edges (circled).}
    \label{fig:epd_cal}
\end{figure}

To calibrate each individual detector chain's gain (e.g., mV/MeV) and offset (keV), several radioisotopes were used as reference inputs. For EPDE, these included Ba-133, Sr-90, and Ru-106. Ru-106 \citep{Eckerman2008} proved to be an invaluable reference isotope for the intended energy range (50 keV to multi-MeV), providing a broad spectrum beta emission up to 4 MeV, alongside several narrow-band gamma ray emissions at 511 keV and 622 keV, which were detectable via their Compton Edge \citep{Felsteiner1974} signatures in each detector chain. The EPDs were configured in a special ground calibration mode with 4096 high resolution energy channels at 4 keV linear steps. The isotope was integrated over a long duration in vacuum and compared with the theoretical Ru-106 spectrum fed into Geant4. Fig. \ref{fig:epd_cal} shows the theoretical Geant result alongside the actual detected counts versus energy channel. The system gains were close to the expected 240 mV/MeV quoted in Amptek A225 literature, but differed sufficiently to justify the test campaign. For EPDI, a similar campaign was carried out using Am-241 \citep{Baranov1963,Terada2016}, using the 59 keV gamma emission and 5.5 MeV alpha particle emission as references.

\begin{figure}[!htb]
    \centering
    \includegraphics[width=\linewidth]{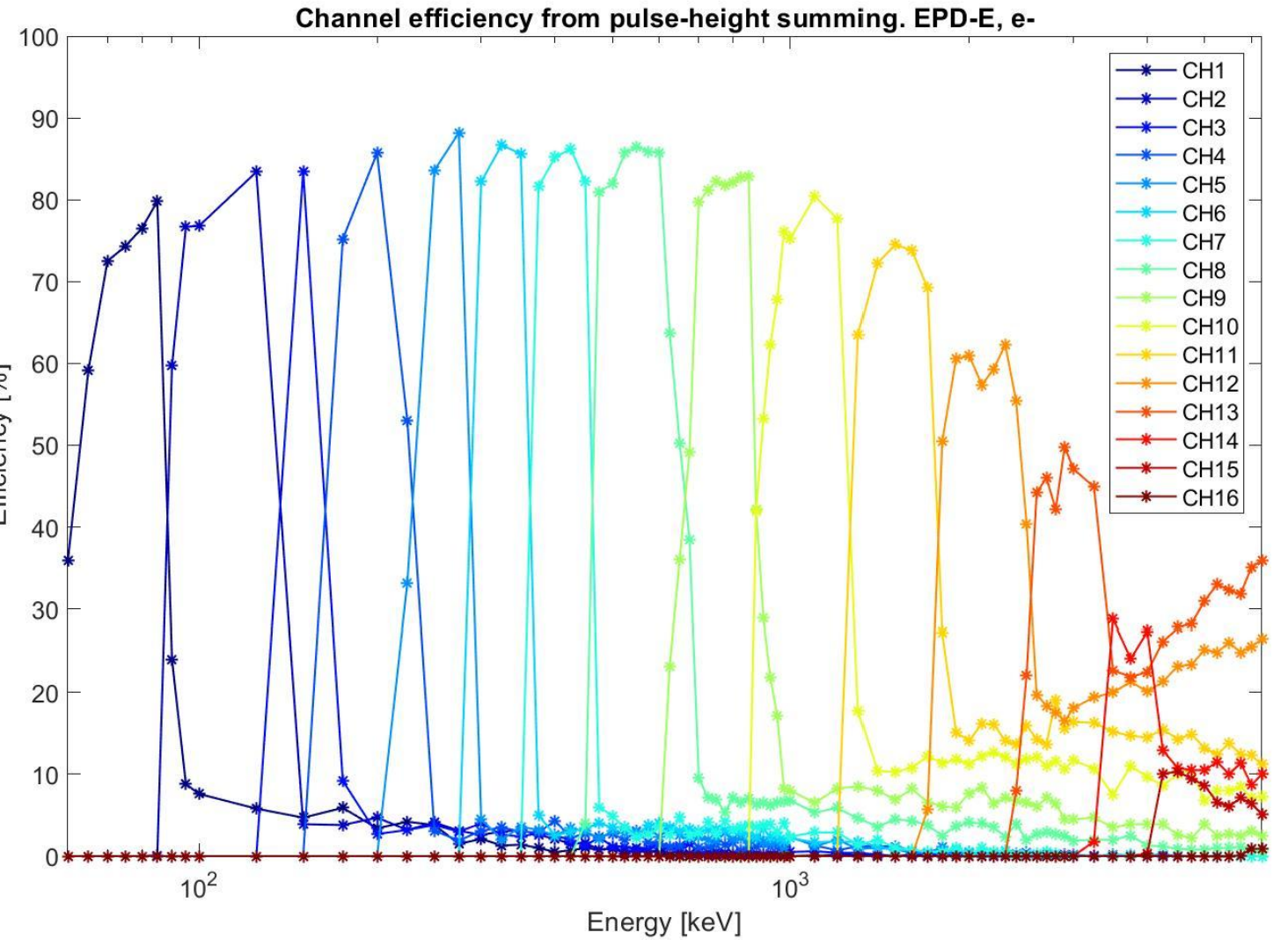}\\
    \includegraphics[width=\linewidth]{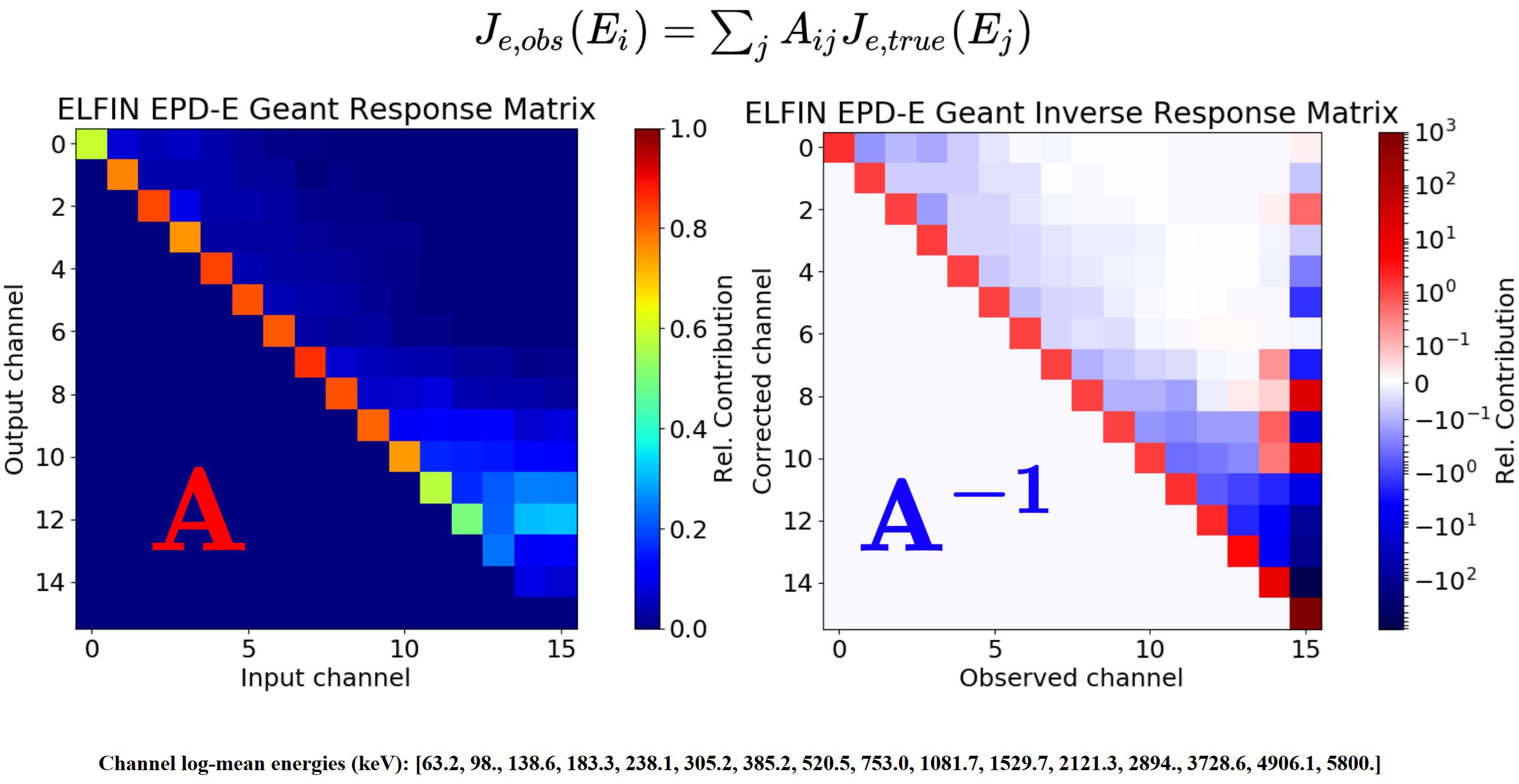}
    \caption{EPDE channel efficiencies computed using Geant4 (top), and resulting channel response matrices (bottom). The channel efficiencies are computed using a large sample of input beams spanning small energy steps (horizontal axis), with the resulting fraction of counts in each of the 16 energy channels shown as colored line plots. The channel response matrix encodes this information by mapping the statistical mixing between the true input energy channel number fluxes $J_{e,true}$ and observed (apparent) $J_{e,obs}$ channels.}
    \label{fig:ch_eff_epde}
\end{figure}

To determine the energy channel efficiencies, defined as the fraction of input beams of a particular energy $E$ depositing energy in each of the 16 log-spaced channels used in flight, Geant4 was used to sweep the electron and proton energy range from 50 keV to 6 MeV. For EPDI, the channel efficiencies are all $\sim$1, due to the total deposition of protons up to 6 MeV in the P1 detector. For EPDE, the efficiencies are more complex, owing to the fact that electrons do not always deposit the same energy per impact within the silicon, and must be treated statistically. The channel response matrix ($\bf A$) computed using Geant shown in Fig. \ref{fig:ch_eff_epde}, which when multiplied to an input energy spectrum, provides the characteristic detection response. The inverse of this matrix (${\bf A}^{-1}$) is correspondingly computed, which can be applied to deconvolve mixed channels. Historically, for simplicity only the diagonal elements of the matrix ($\bf A$) have been used as dividing factors in calibration codes. 

The EPDE is seen to provide high efficiency (low channel mixing) up to several MeV, after which the sensitivity is reduced but remains non-zero. In part for this reason, channel 16 is purposed on-board as an energy-integral channel. The EPDI is technically sensitive up to 6 MeV as well; however, due to its much smaller geometric factor, it typically did not see particle counts above 3 MeV in any particular spin sector.

\subsection{Noise assessment, filtering, and thresholding}
\begin{figure}[!htb]
    \centering
    \includegraphics[width=\linewidth]{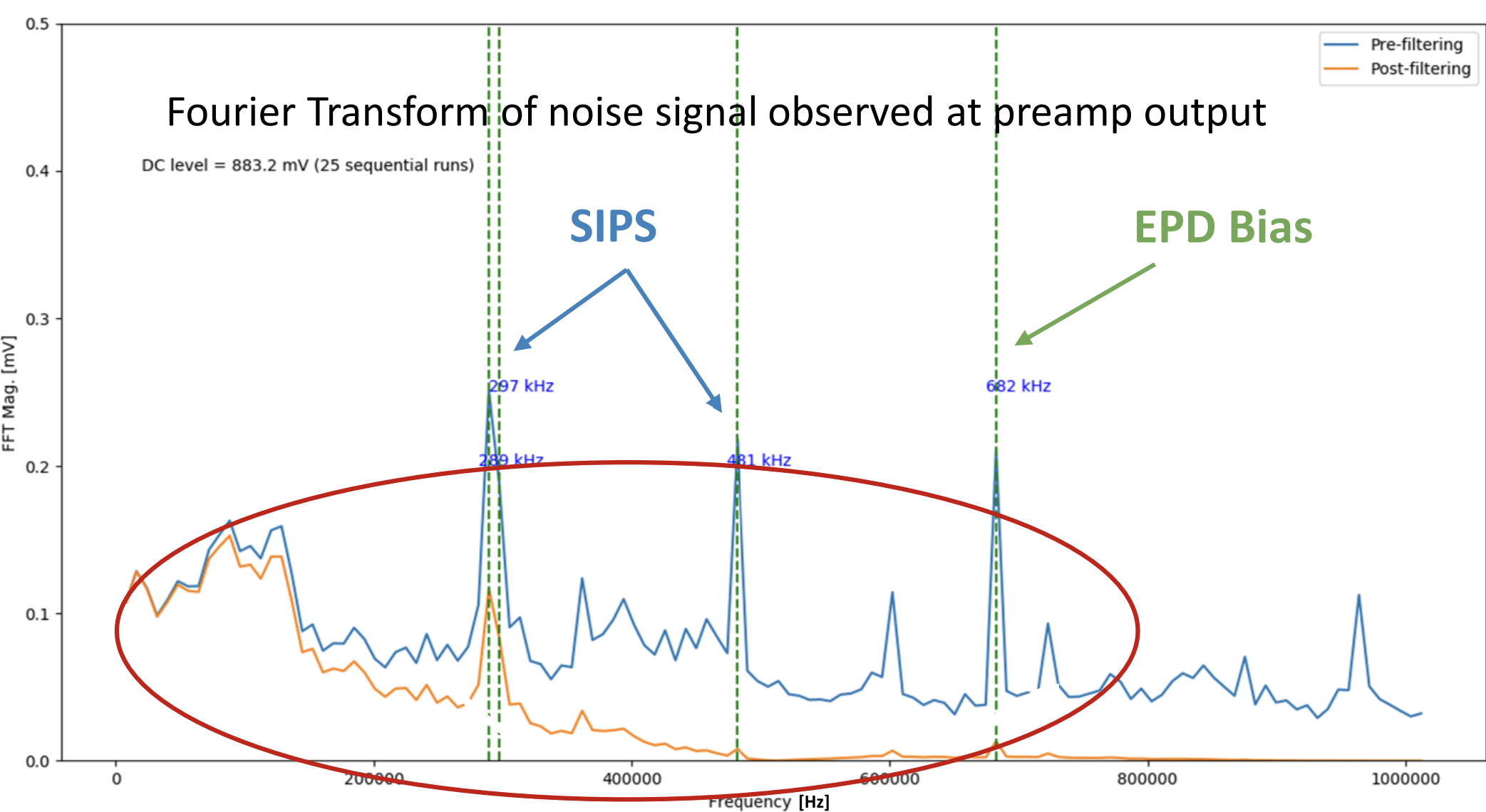}\\
    \includegraphics[width=\linewidth]{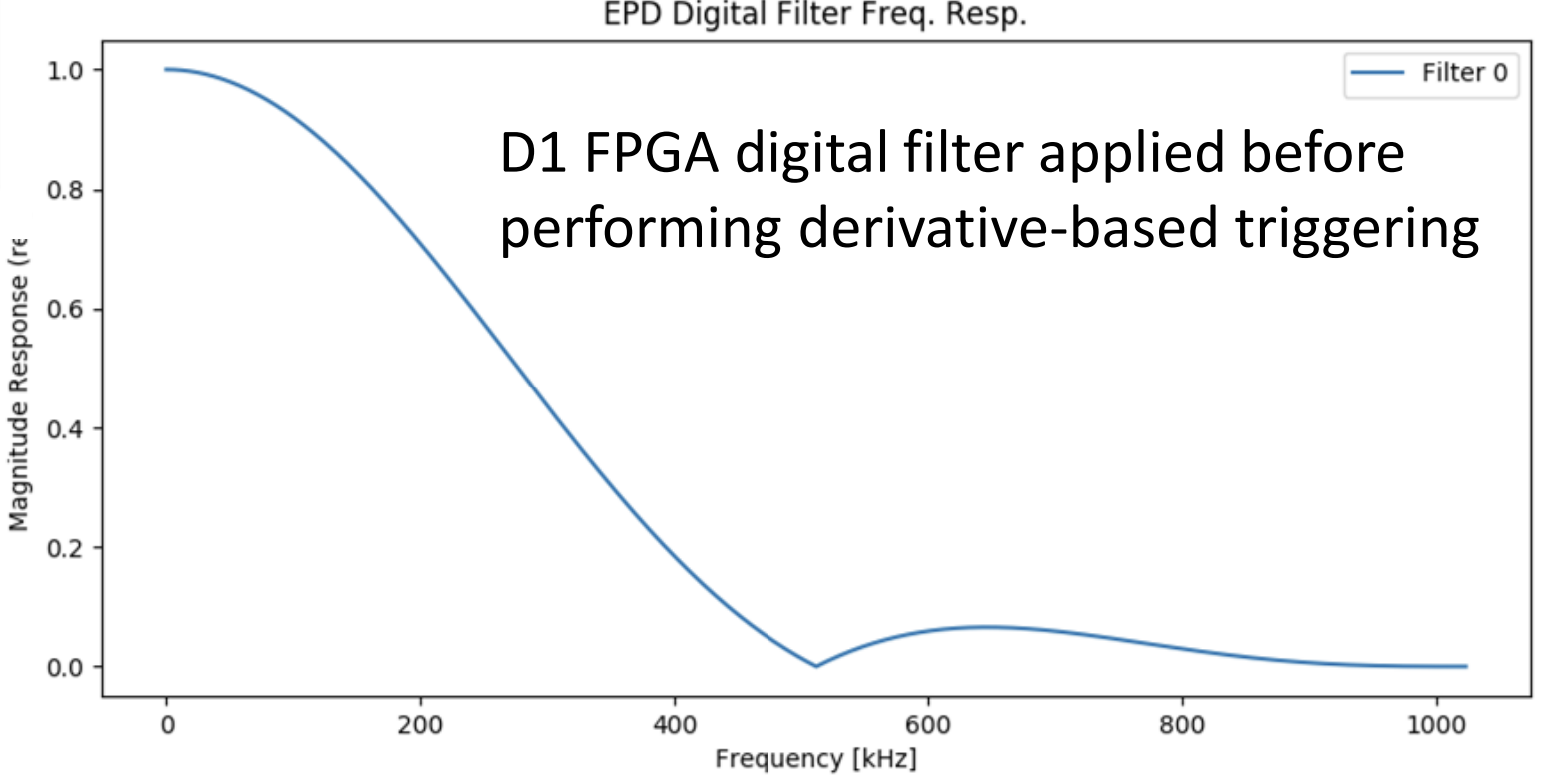}
    \caption{Fourier transform magnitude of internal noise sources appearing at pre-amplifier inputs (top), and normalized digital finite-impulse response (FIR) filtering applied to the signal (bottom).}
    \label{fig:spectral}
\end{figure}

In order to minimize and reject spurious counts due to noise signals, several approaches were taken. By design, the detector leads were physically placed and oriented to minimize cross-talk. During design and testing, the intrinsic frequency response of the overall electronics chain was also evaluated using Fourier analysis. This was achieved by taking FFTs of the ADC output time series of each detector chain under various conditions. An example is shown in Fig. \ref{fig:spectral}, where the frequency contributions from various system elements (e.g., SIPS driver, bias clock) are readily apparent. Where possible, these frequencies were adjusted by trial and error to avoid the critical frequencies of the preamp pulse shaper signal. A digital FIR filter was also implemented in the D1 FPGA to reject frequencies out of this band, resulting in significant attenuation of internal interference/cross-talk.

\begin{figure}[!htb]
    \centering
    \includegraphics[width=\linewidth]{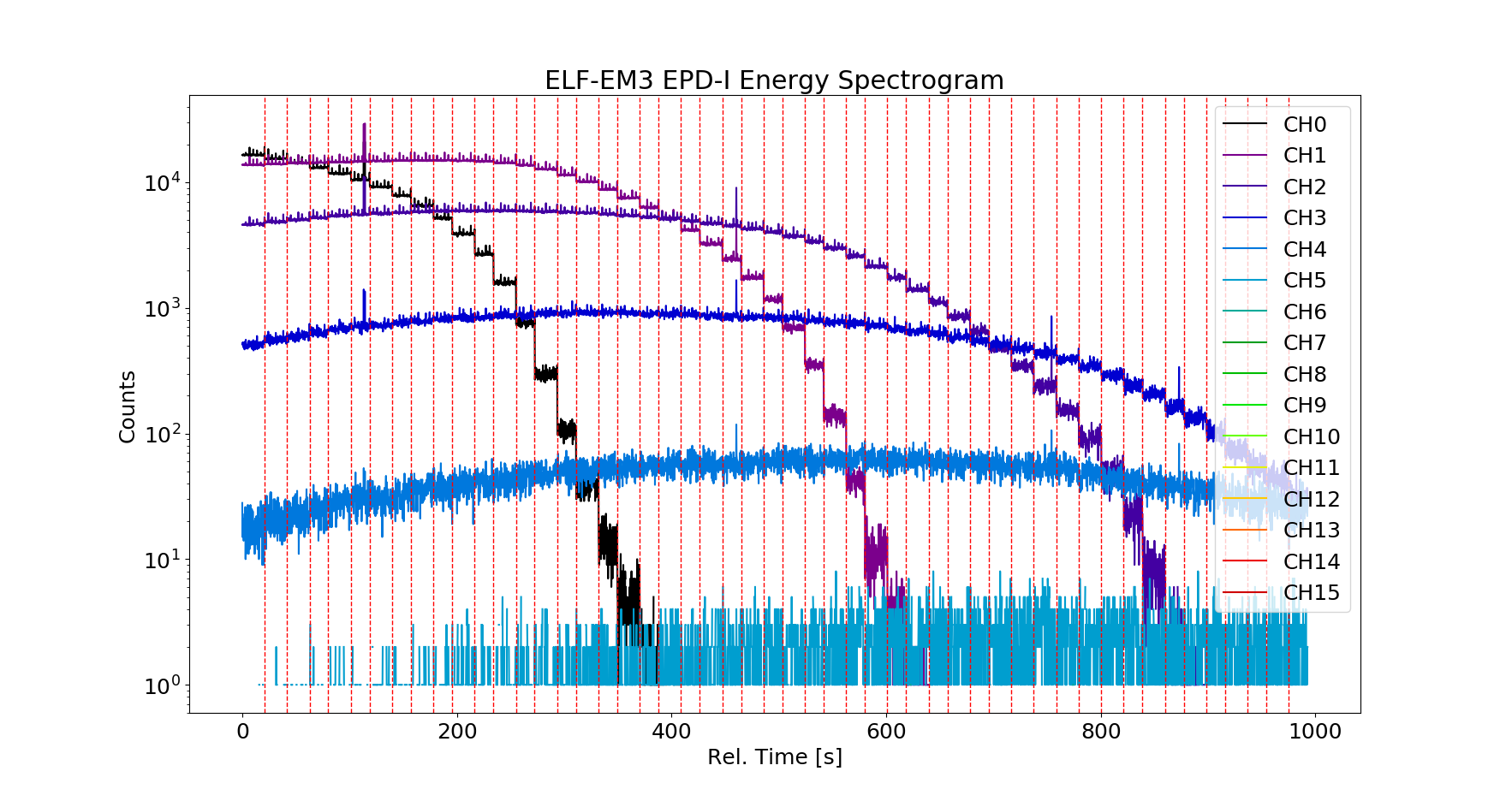}\\
    \includegraphics[width=\linewidth]{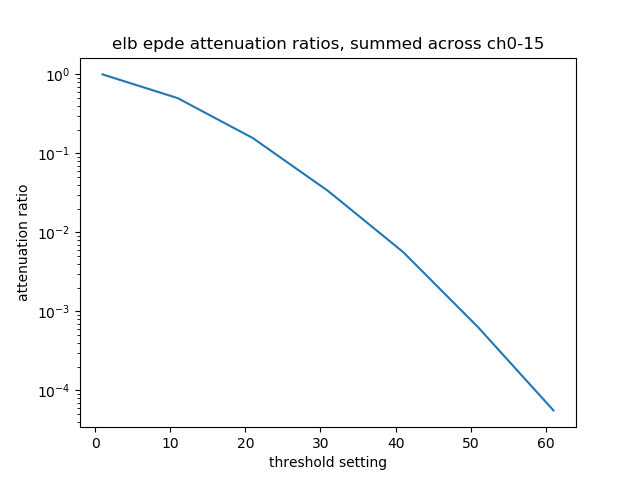}
    \caption{Noise thresholding procedure, showing the resulting noise count rates in a special calibration set of low-energy channels as a function of increasing differential start trigger threshold (vertical red bar increments over time) on the EM3 unit (top), and resulting integral noise count rates versus each threshold setting in the flight-model EPD on ELFIN-B (bottom). As the threshold is increased, spurious noise counts are observed to decrease.}
    \label{fig:thresholds}
\end{figure}

Although these mitigation measures significantly reduced the noise level, they did not completely eliminate it. For this reason, as described above, detector signal chain thresholds were employed to automatically reject pulses below configurable levels. To find the appropriate setpoints in flight, EPD was operated at the geomagnetic equator where fluxes of energetic electrons and protons are typically below the sensitivity of the instrument. In order to distinguish noise from actual particle fluxes at this location, the data were required to possess a comparable spectrum to that measured in ground testing, and which exhibit negligible variation in the spectrum as the satellites spin. The resulting observations are therefore taken to be the intrinsic noise spectrum of the instrument. To find the appropriate threshold setpoint for each isolated detector, the threshold values were incremented sequentially until finding a threshold such that the noise count rate production dropped to below $10^{-4}$ of its unthresholded level, corresponding to less than 1 noise count per spin sector on average. An example of this procedure on the E1 detector is shown in Fig. \ref{fig:thresholds}, using the engineering model (EM3) and corresponding flight results on the FM-2 EPD (ELFIN-B). In these tests, the energy channel bounds were also configured to a special low-energy mode, allowing for fine-scale resolution of the noise spectrum. Based on this procedure, a single set of threshold values was used throughout the mission, with corresponding cutoff energies below 50 keV.

\section{Data products}
\subsection{Product descriptions}
We now describe the nominal EPD science data products, which can be straightforwardly loaded and processed using the SPEDAS software framework \citep{Angelopoulos2019}. EPD data is released in three product levels: ``raw''/count rate (L0) data; calibrated number- and energy-flux spectrogram (L1) data; and calibrated pitch-angle energy spectrogram (L2) data. The names and descriptions of each data product, as well as key auxiliary quantities, are tabulated in Table \ref{tab:data_products}.

To demonstrate loading and interpreting the data products, as well as showcase the previously unreported EPDI capabilities, we consider an example ELFIN-A Outer Belt Observation (OBO) science collection on 2022-08-20 from 01:20 to 01:26 UTC. The satellite was on a south-descending trajectory, beginning at $\sim$22 MLT in the outer radiation belt (electrons) and the ring current (ions). Over several minutes, it transitions into the plasma sheet, ending in the polar cap around 21 MLT. The L0, L1, and L2 products in this collection will be explored, along with configuration options. The EPD data for this and any other event can be loaded in several ways:
\begin{enumerate}
    \item SPEDAS (IDL), using the \verb|elf_load_epd| routine (L0 and L1 data), followed by the \verb|elf_getspec| routine (L2 pitch-angle products)
    \item PySPEDAS (python), using the \verb|pyspedas.projects.elfin.epd| routine (all data levels)
    \item Custom user software, by manually loading the CDF data files stored on the ELFIN data servers
\end{enumerate}

\subsection{L0 data and processing}
\begin{figure}[!htb]
    \centering
    \includegraphics[width=\linewidth]{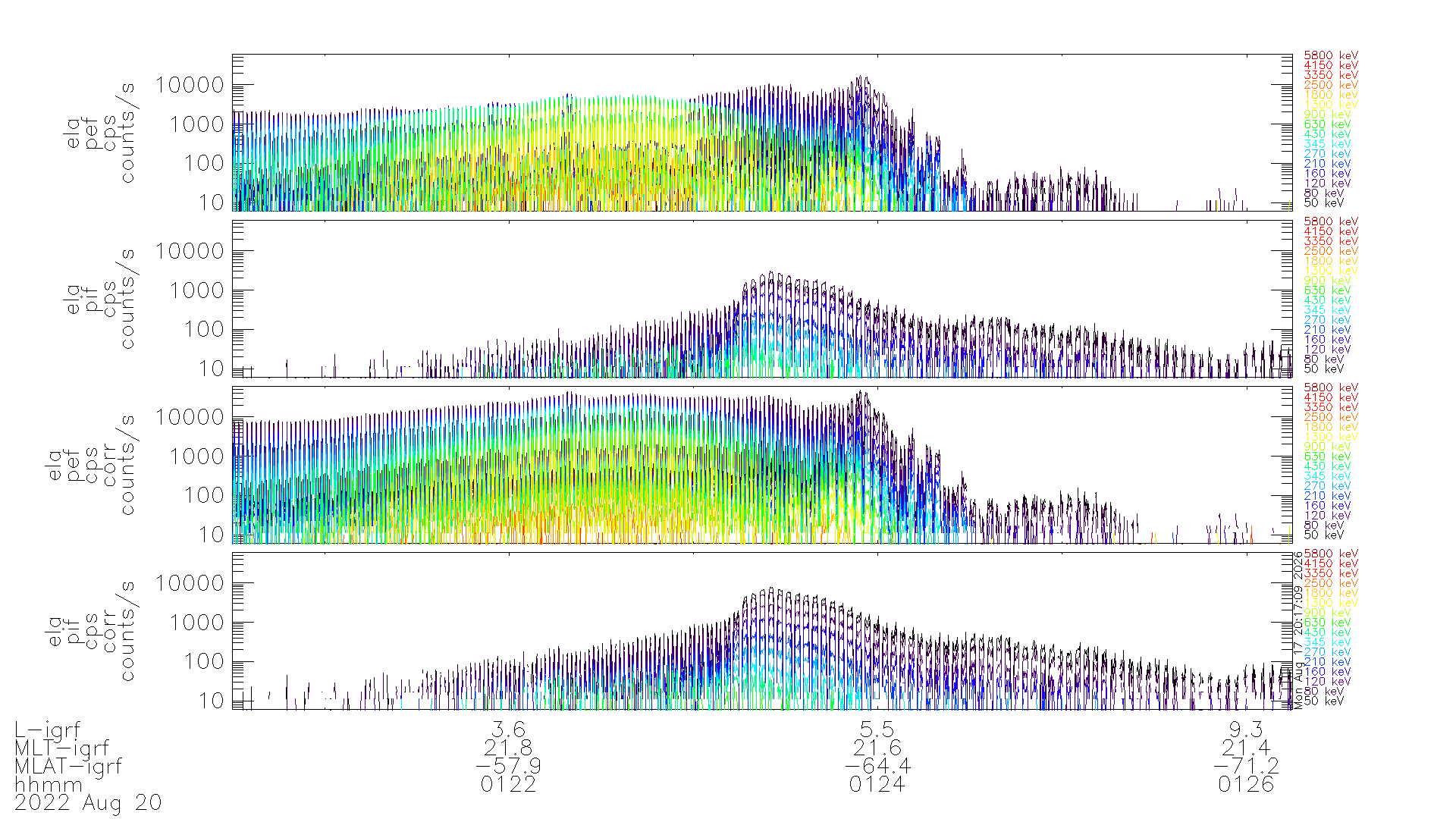}\\
    \includegraphics[width=\linewidth]{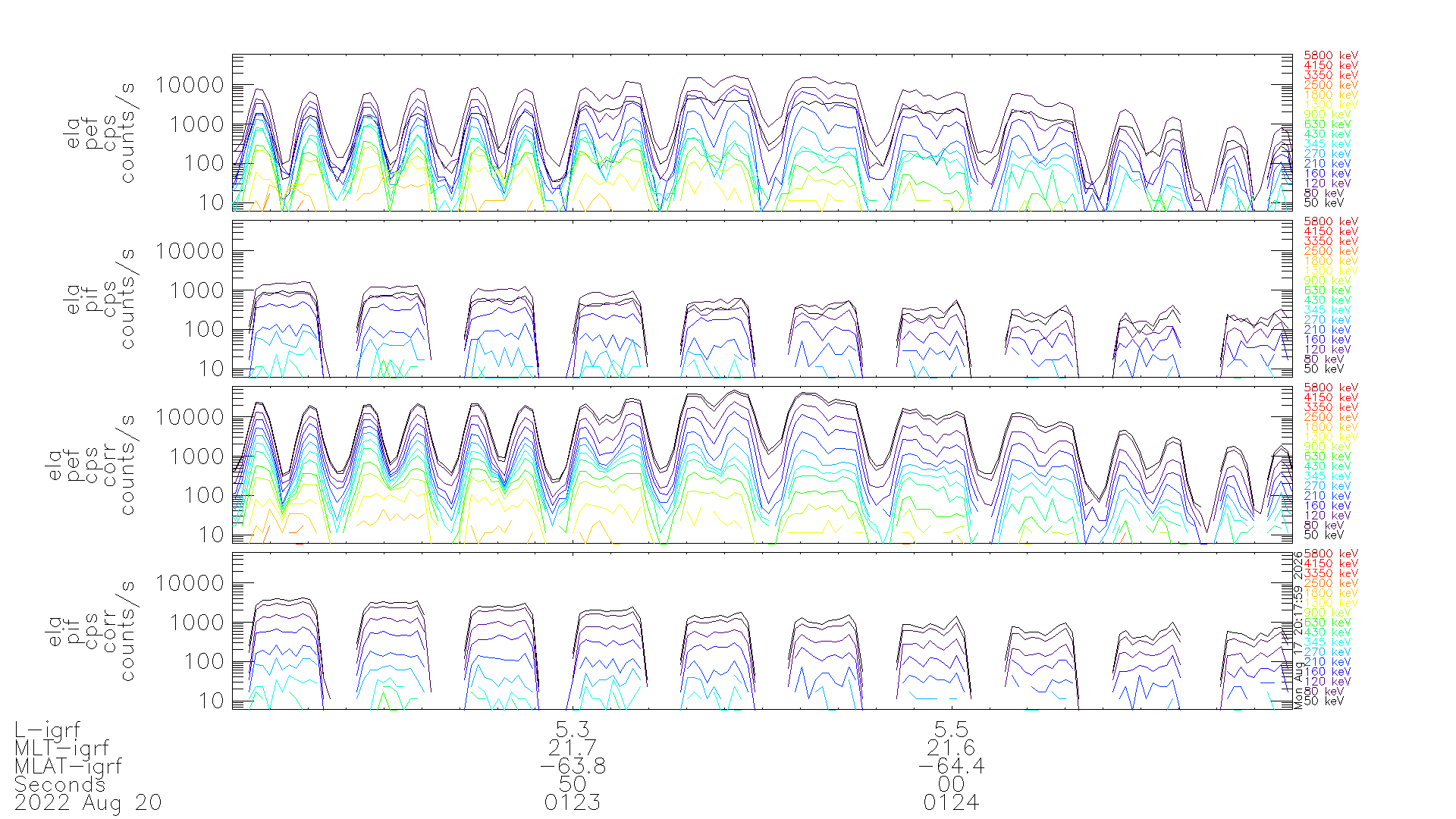}
    \caption{EPD L0 data example of the counts per second (cps) product captured by ELFIN-A on 2022-08-20 from 0120-0126 UT for electrons and ions with (first and second panels, top) and without (third and fourth, top) paralyzable detector correction applied, respectively. Each line plot shows a different energy channel, annoted in color on the right-hand side. Bottom four panels show these data in a zoomed-in time interval containing roughly 10 spin periods, near the native instrument time resolution.}
    \label{fig:L0_data}
\end{figure}

We begin by examining L0 (least-processed) data, including two distinct time series products per particle species (e and i): (1) ``raw'' counts per spin sector in each of the 16 energy channels; and (2) counts per second (``cps'') in each spin sector and energy channel. The cps product is computed by converting each sector into its equivalent time width, found by subdividing the satellite spin period (typically 2.81 sec) by 16 sectors. The time-varying spin period used in the calculation is provided as an auxiliary product. Using IDL SPEDAS, Fig. \ref{fig:L0_data} (top) Panels 1 and 3 shows count-rate time series, with a separate colored trace for each energy channel, for electrons and ions, respectively. Rapid variations in count level correspond to spin motion of the satellite. Over this timescale, the EPD senses a conical cut of particle distributions relative to the local background geomagnetic field, including parallel and perpendicular look directions, which often exhibit flux extrema relative to other directions. Panels 1 and 3 of Fig. \ref{fig:L0_data} (bottom) show a zoomed-in time window from the collection, containing several spin period's worth of cps data. Typical variations in count rates at different points within each period can be observed, which in this interval correspond to an anisotropic ``trapped'' population of electrons and an isotropic ion population (excluding the back-scattered direction). 

The L0 cps data do not inherently account for the ``dead time'' behavior of the EPD signal chain, in which some counts are lost due to previous in-progress impact events. In order to correct for this effect, the particular signal chain and coincidence logic can be modeled as a paralyzable system \citep{Usman2018}, in which counts at higher energies tend to override in-progress counts at lower energies. Consequently, the count rate dead time effects in any particular energy channel depend not only on its own value, but also on the sum of all counts in channels above it. Under the assumption of Poisson statistics, this can be stated as follows:
\begin{equation}
    C_{det,\geq i} = C_{true,i}e^{-C_{true,i}/C_\tau}
\end{equation}
where $C_{det,\geq i}$ is the detected count rate summed over including and above the $i^{th}$ energy channel, $C_{true,i}$ is the true input count rate for channel $i$ alone, and $C_\tau = (2.4$ $\mu$s$)^{-1}$ is the reciprocal of the characteristic ``dead time'' of the system, limited by the rise time of the preamp pulse shaper. 

\begin{figure}[!htb]
    \centering
    \includegraphics[width=\linewidth]{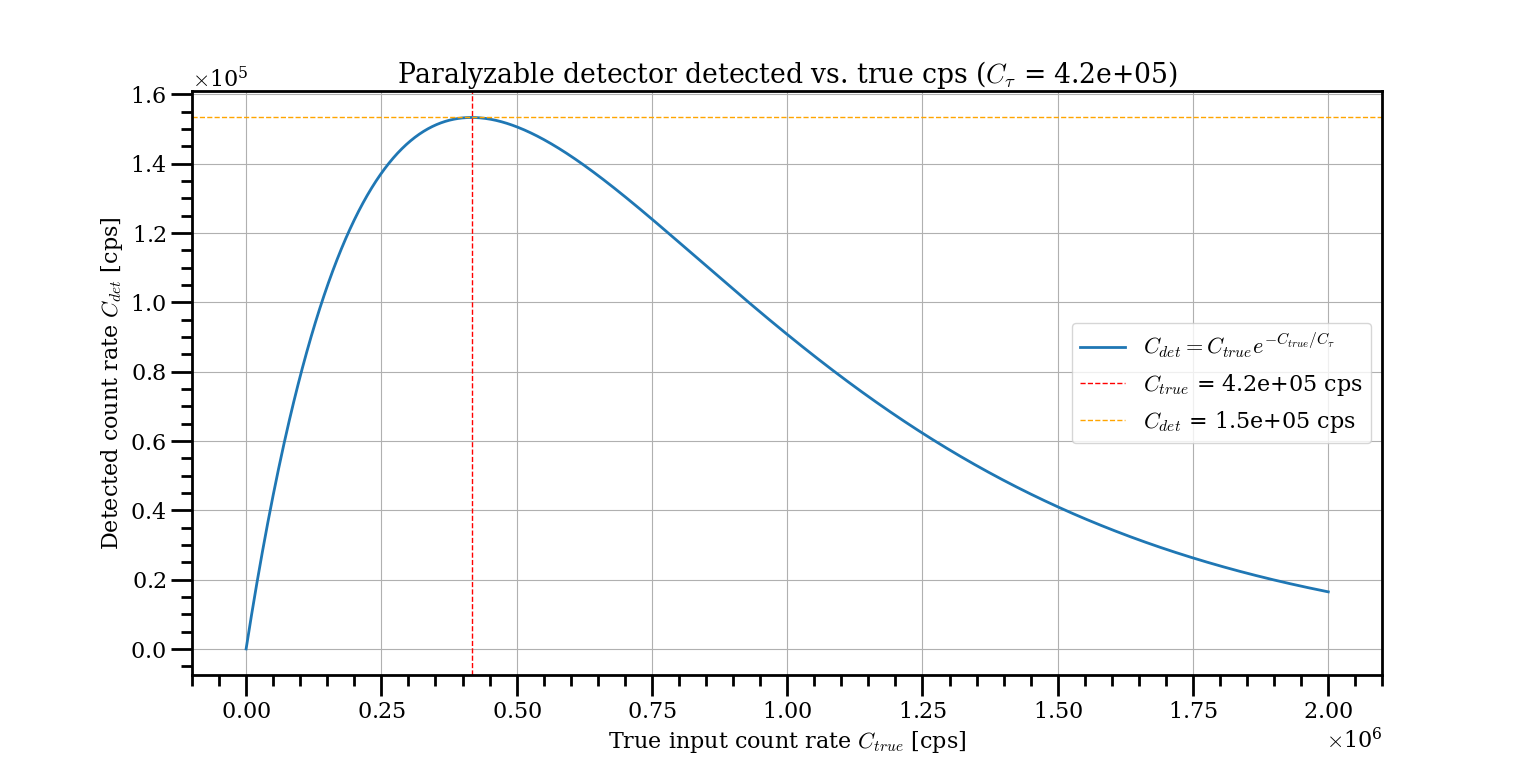}
    \caption{EPD detected versus true input counting rate, modeled as a paralyzable system limited by finite pulse-shaper rise time and differential triggering (single idealized energy channel). The spectral response as a function of multiple energy channels is described by Eqn. 1.}
    \label{fig:coi_logic_paralyzable}
\end{figure}

An example of this relationship is shown in Fig. \ref{fig:coi_logic_paralyzable}, where in this simplified case, the detector channel is the highest energy receiving non-zero counts (i.e. $C_{det,i}=C_{det,\geq i}$). It can be seen that the detected count rate is double-valued about the saturation level, reaching a maximum detected rate of $1/e$ times the true count rate. In practice, this true rate is found by inverting Eqn. 1 using the Lambert W function \citep{Corless1996}, nominally taking the branch cut with absolute count rate below the dead time as representative. For extremely high count rate events (e.g. storms), care should be taken in evaluating the validity of this branch cut.

The paralyzable detector correction can be applied using the \verb|apply_deadtime_corr| keyword to the \verb|elf_load_epd| routine in IDL SPEDAS, and is applied automatically for L1/L2 data loaded via CDF files (PySPEDAS and custom user software). Panels 2 and 4 of Fig. \ref{fig:L0_data} (top, bottom) show the result of applying the correction to the cps data in Panels 1 and 3. It can be observed that the count rates are boosted -- especially at lower energies, which experience the largest correction, owing to the count summing in higher energy channels. Count rates in higher energy channels can also be significantly boosted by the correction during events where summed count rates approach the reciprocal of the system dead time, such as storms and substorms. The spectra are also observed to become monotonically decreasing in energy, which is an imposed property of the correction algorithm. Care should be taken in choosing whether to apply the correction in circumstances where distributions are suspected to possess atypical (e.g., unstable/non-monotonic) distribution functions on physical grounds. For channels with zero counts, the correction is not applied, even if channels above or below it have non-zero counts. This allows narrow energy band structures to be captured in the data (e.g. wave-particle interactions or SEP events) without artificially enhancing the counts in channels below it. 

\subsection{L1 data and processing}
Level 1 EPD data are comprised of a transformed version of the L0 cps product into differential-directional particle fluxes, resulting in either number flux (``nflux'') or energy flux (``eflux''). This transformation is performed automatically as part of the IDL SPEDAS load routine, or can be loaded directly from EPD CDFs (PySPEDAS and custom software). 

For each detector head, the conversion from cps $C_n$ to number flux $J^N_n$ as a function of the $n^{th}$ energy channel of lower bound $E_n$ is performed by applying several calibration factors: (1) detector head geometric factor $G$ (units of cm$^2$ str); (2) differential energy channel width taken as the difference of adjacent lower channel bounds $dE_n=E_{n+1}-E_n$ (units of MeV); and (3) energy-dependent channel efficiencies $A_n$ (dimensionless, $\leq$1):
\begin{equation}
    J^N_n = \frac{C_n}{G\cdot dE_n\cdot A_n}
\end{equation}
From the number flux $J_n$, the equivalent energy flux $J^E_n$ is computed as:
\begin{equation}
    J^E_n = \bar{E}_n\cdot J^N_n
\end{equation}
where $\bar{E}_n$ is the geometric mean of the particular energy channel in keV (e.g. 63.2 keV for the 50-80 keV channel). The 16th channel is treated specially as integral flux above its minimum energy, with an \emph{a priori} energy width $dE=1$ MeV presumed. This width was based on the maximum energy range expected to be encountered with sufficiently high number fluxes as to be detectable by the combination of ELFIN's orbit and EPD sensitivity.

\begin{figure}[!htb]
    \centering
    \includegraphics[width=\linewidth]{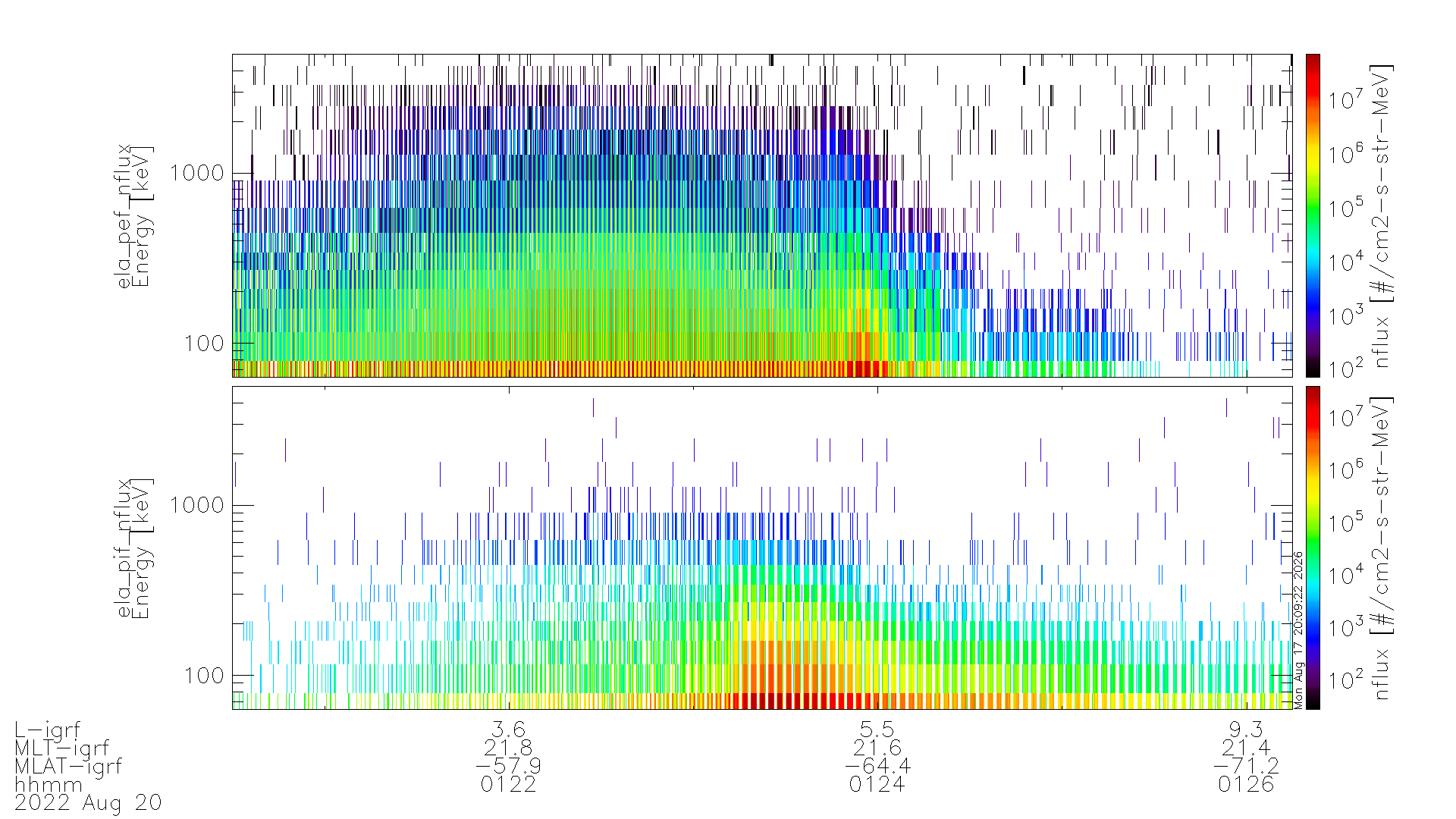}\\
    \includegraphics[width=\linewidth]{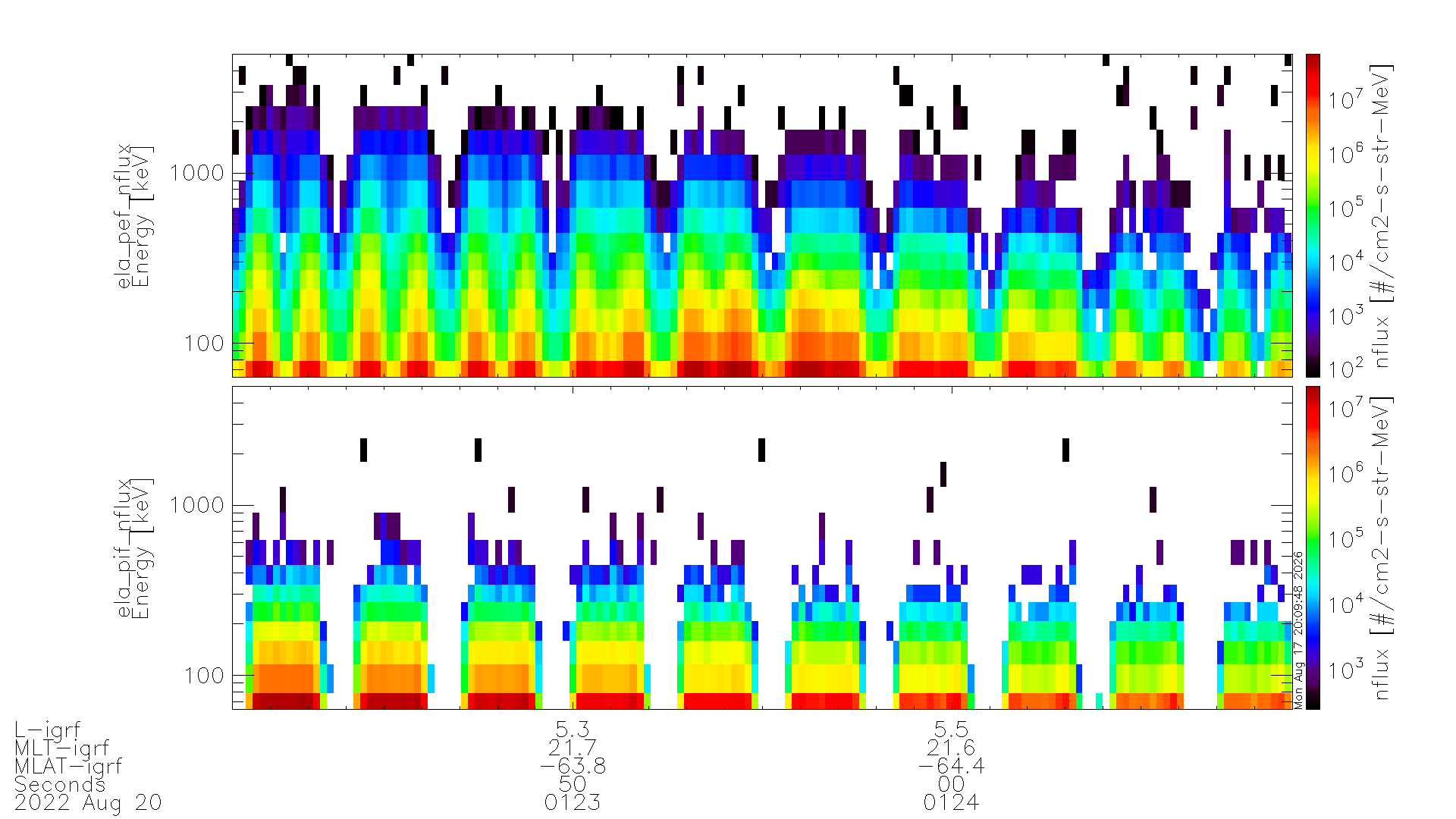}
    \caption{EPD L1 data example of the differential-directional number flux (nflux) energy spectrum product corresponding to the same interval in Fig. \ref{fig:L0_data}. The vertical axes are now energy, with the spectral axis in units of 1/cm$^2$ s str MeV.}
    \label{fig:L1_data}
\end{figure}

Fig. \ref{fig:L1_data} shows an example stack plot of electron and ion number flux data. Panels 1 and 2 (top, bottom) show the equivalent dead time-corrected nflux derived from the L0 cps data shown in Panels 2 and 4 (top, bottom) of Fig. \ref{fig:L0_data}. The values span a very wide dynamic range ($\leq10^2$ to $\geq10^8$ 1/cm$^2$ s str MeV), demonstrating the versatility of the ELFIN EPD in a range of conditions.

We note that the default channel efficiency calibration ($A_n$ value) presumes that each energy channel is independent of the others (negligible mixing), which should be interpreted with care for electrons above $\sim$3 MeV. Additionally, when highly non-monotonic energy spectra are present (without dead time correction applied), EPDE and EPDI data should be compared directly to check for potential cross-contamination. These effects are negligible in most events but can be significant during extreme cases, such as Solar Energetic Particle (SEP) events, where protons tend to be sufficiently energetic and high flux to penetrate the front foil.

\subsection{L2 data and processing}
\begin{figure}[!htb]
    \centering
    \includegraphics[width=\linewidth]{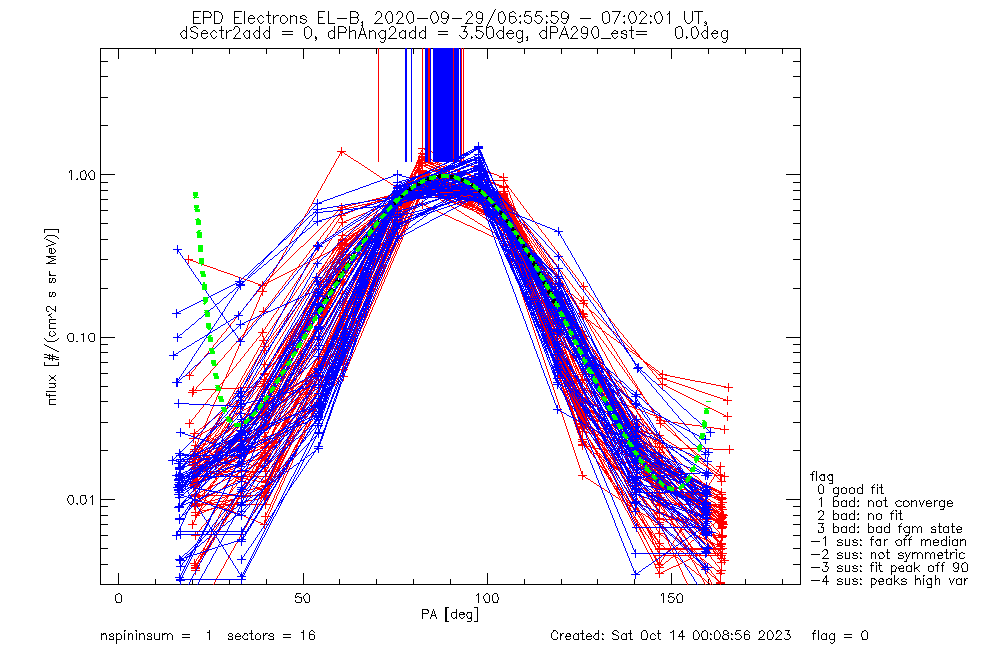}
    \caption{Example superposed pitch-angle distributions and fit from consecutive spin periods (one per 2.8 s) in the roughly 2 minute central window of the ELFIN-B collection on 2020-09-29 at 0656 UT. The procedure operating over all spin periods converged to the curve in dashed green color, resulting in a good fit with pitch-angle correction (phase delay) of 3.5$^\circ$ for this particular interval.}
    \label{fig:pa_fits}
\end{figure}

To form the L2 data, the L1 nflux or eflux products are converted into pitch-angle spectrograms sharing the same physical units. This is achieved using a procedure developed to find the best-fit phase angle (delay) to a polynomial spline over a subset of spin periods within a collection interval (6-7 minutes), based on particle flux peaking symmetrically about $90^\circ$. An example is shown in Fig. \ref{fig:pa_fits}. The quality of such fits is automatically classified based on their apparent goodness. This procedure is usually successful, but care should be taken in interpreting the data to ensure that the phase offset is physically plausible. Additionally, if the data do not possess a subset of spin periods with symmetric flux peaking, the algorithm will not succeed. Such cases must be analyzed manually to determine the best-fit phase correction.

With the phase delay correction known, individual pitch-angles corresponding to each spin sector are assigned from the interpolated best-fit solution. This procedure is provided as part of the IDL SPEDAS framework via the \verb|elf_getspec| routine, with the output also being stored in the L2 CDF files. The resulting data product takes the form of a pitch-angle distribution as a function of energy at either full-spin or half-spin time resolution. The two spin resolutions differ by whether the exact pitch-angles in the ascending versus descending portion of the spin period are considered separate elements at higher time resolution (e.g. for studying micro-bursts), or whether they are added/averaged together to decrease counting uncertainty. We note that half-spin data cannot alone distinguish whether differences in flux measurements at conjugate pitch-angles are due to satellite track motion or to the start time of a fast timescale process (e.g. micro-bursts). 

\begin{figure}[!htb]
    \centering
    \includegraphics[width=\linewidth]{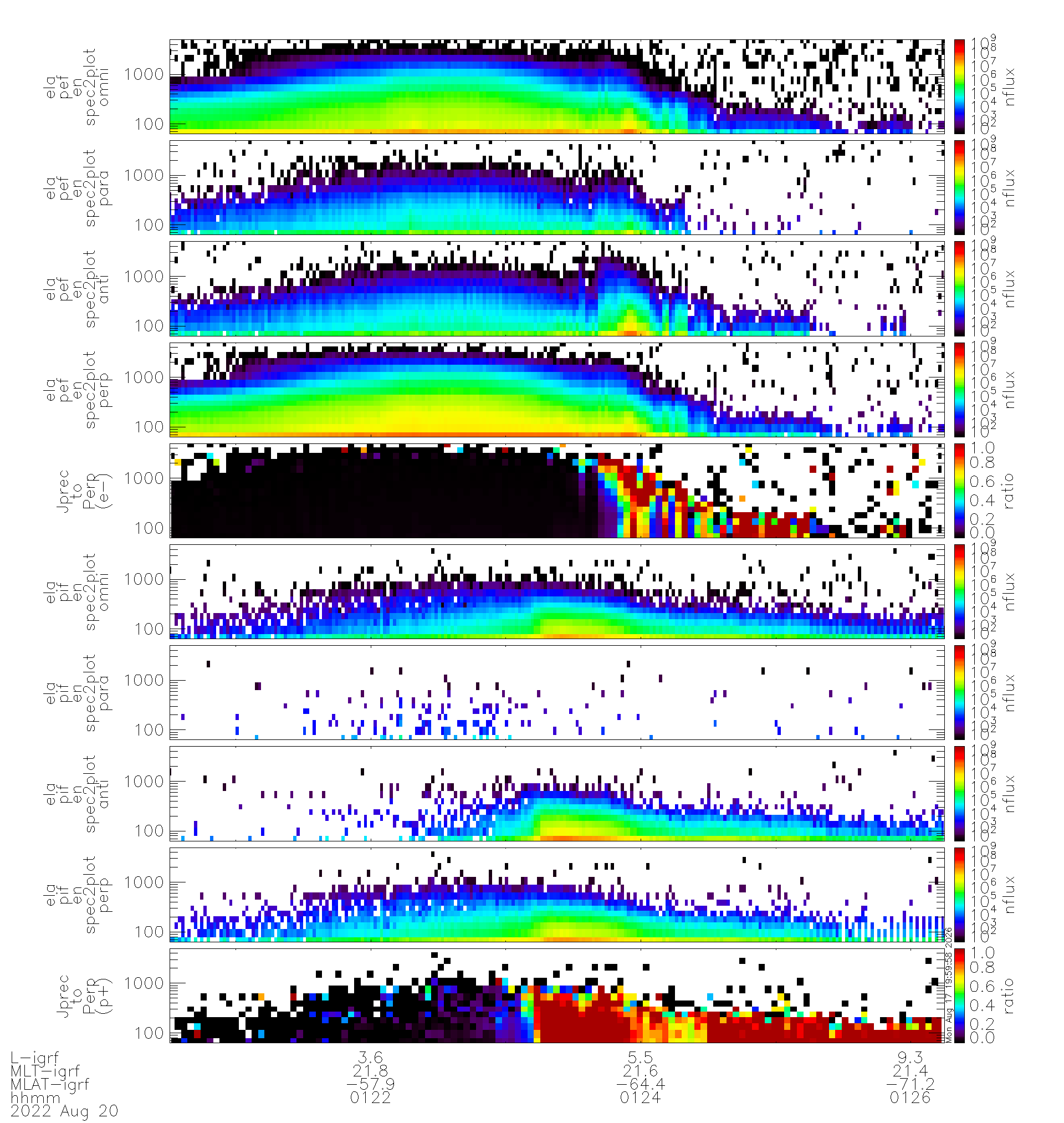}
    \caption{EPD L2 data example of energy spectra at selected pitch-angles corresponding to the same time interval in Fig. \ref{fig:L1_data} (top) for electrons (panels 1-5) and ions (panels 6-10). ``omni'' refers to omnidirectional, ``perp'' to perpendicular, ``prec'' to precipitating, and ``back'' to back-scattering pitch-angle ranges, respectively.}
    \label{fig:L2_data_specplots}
\end{figure}

\begin{figure}[!htb]
    \centering
    \includegraphics[width=\linewidth]{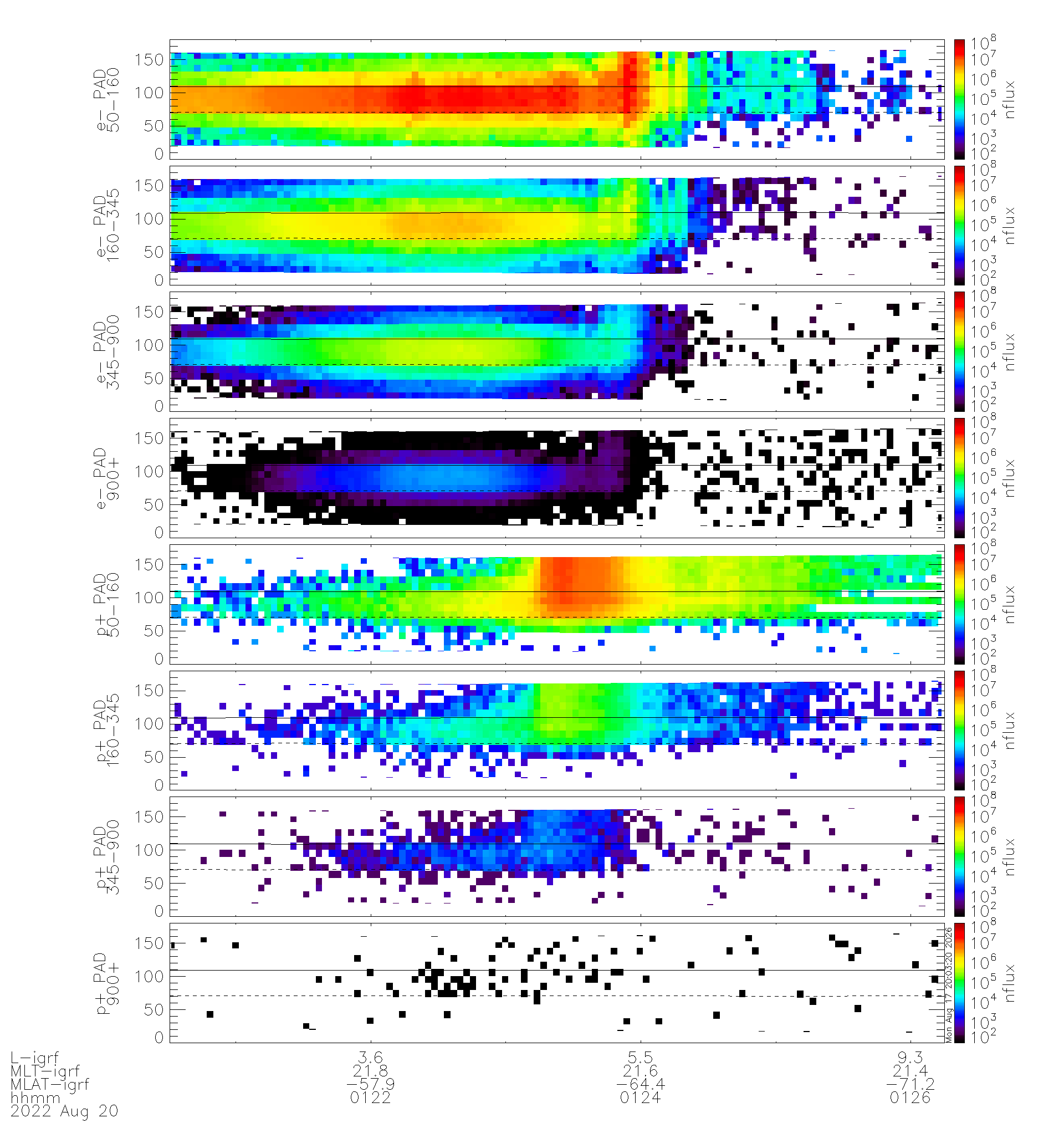}
    \caption{EPD L2 data example of pitch-angle spectrograms for selected energy ranges corresponding to the same time interval in Fig. \ref{fig:L1_data} (top) for electrons (panels 1-4) and ions (panels 5-8). Energy binning ranges are indicated in the vertical captions.}
    \label{fig:L2_data_pads}
\end{figure}

The L2 pitch-angle spectra form 3-d arrays in time, energy, and pitch-angle, making them challenging to visualize simultaneously along all axes. Instead, a common approach is to reduce the data set in one of two ways: (1) collapsing the pitch-angle dimension into four physically meaningful ranges: omnidirectional (``omni''), perpendicular (``perp''), precipitating (``prec''), and back-going (``back''); and (2), to reduce the energy dimension into four ranges, by default equal to 50-160 keV (``ch1''), 160-345 keV (``ch2''), 345-900 keV (``ch3''), and $\geq$900 keV (``ch4''). Examples of both reduction schemes based on L1 nflux data are shown in Fig. \ref{fig:L2_data_specplots} and Fig. \ref{fig:L2_data_pads}, respectively. The pitch-angle distributions are presumed to be gyro-tropic.

In case (1) (Fig. \ref{fig:L2_data_specplots}), omnidirectional refers to averaging across the entire spin period. Perpendicular refers to counts at pitch-angles outside of the bounce loss-cone (LC) and anti-loss-cone angles, which are computed as the pitch-angle for which any particle can reach 100 km altitude (presumed lost to the atmosphere). Precipitating refers to particles with pitch-angles colliding with the local downward atmosphere in the hemisphere where data is collected (angle less than LC or greater than 180$^\circ$-LC). Back-going refers to particles whose pitch-angles correspond to precipitation in the conjugate hemisphere of the particle bounce motion. The ratio of precipitating to perp flux is a direct measure of loss-cone filling, which approximates pitch-angle isotropy. This ratio especially useful in identifying physical phenomena, later highlighted in two published examples. 

In case (2) (Fig. \ref{fig:L2_data_pads}), the energy dependence of the pitch-angle distributions can be seen. This type of stack plot is useful for identifying energy-dependent physical phenomena, such as pitch-angle scattering from plasma waves (e.g. EMIC) and magnetic field-line curvature (FLCS), resulting in energy-latitude dispersed isotropy boundaries.

We note that the angular smearing in the field of view caused by satellite rotation can have a minor effect on the angular pitch-angle resolution. In practice, however, the 100 km particle bounce loss-cone remains resolved except for a fractional uncertainty in its edges. In general, the central loss cone receives the most integration time regardless of spin phase.

\subsection{Filtering by counting uncertainty}
\begin{figure}[!htb]
    \centering
    \includegraphics[width=\linewidth]{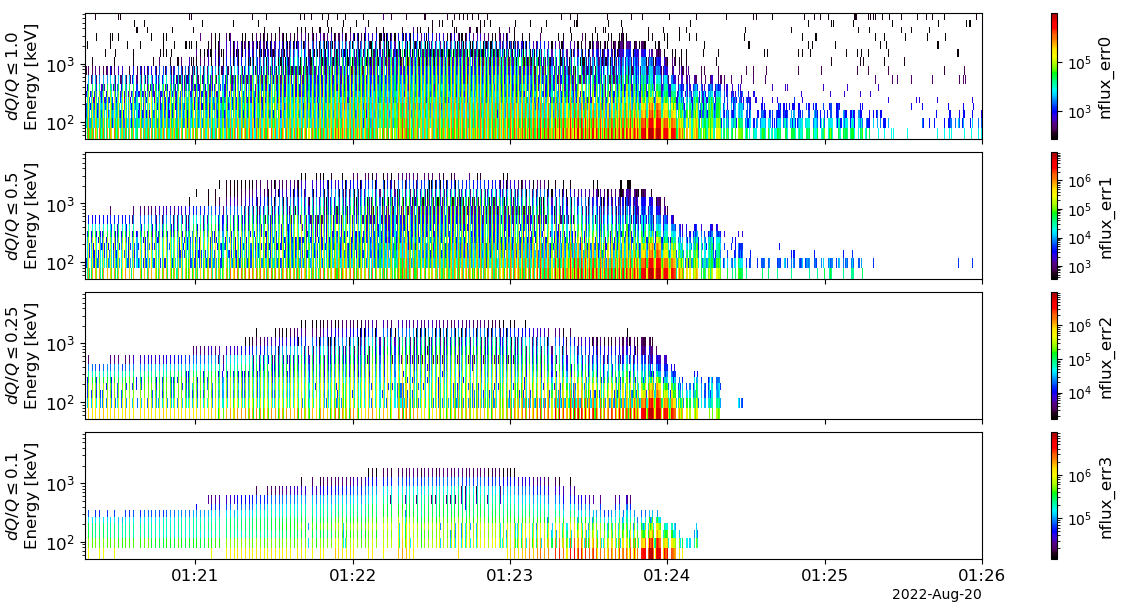}
    \caption{Example of data filtering by $dQ/Q$ error (counting uncertainty). In this example L1 nflux data is used, but the method applies to any product level/type. The case of $dQ/Q=1$ corresponds to unfiltered data.}
    \label{fig:dQovQ}
\end{figure}

A common task in analyzing particle data is to filter it based on the uncertainty in the measured statistical distribution. Under the assumption of Poisson count statistics \citep{Matsunawa2014}, the standard error in the measured counts of each energy channel can be estimated directly from the raw L0 product, where the variance is equal to the total measured counts. The standard deviation is therefore equal to the square root of the measured counts and can be normalized to provide a convenient time-dependent error metric. This value is referred to as $dQ/Q$ for any quantity $Q$ derived from the raw data in the $n^{th}$ energy channel, and is defined as:
\begin{equation}
    \left(\frac{dQ_n}{Q_n}\right)(t) = \frac{1}{\sqrt{R_n(t)}}
\end{equation}
where $R_n(t)$ are the raw count totals accumulated in each individual spin sector at time $t$ in energy channel $n$. For non-zero count rates, the worst-case counting error corresponds to $dQ/Q=1$ (the one-count level), and tends toward zero error as count rates increase. Data can therefore be filtered based on this value, as shown in Fig. \ref{fig:dQovQ}, using L1 nflux data in PySPEDAS as an example. The top panel provides no filtering whatsoever ($dQ/Q \leq 1$), and therefore includes isolated spurious noise counts, and occasional cosmic ray impacts. The second panel introduces an error filtering constraint of $dQ/Q \leq 0.5$, equivalent to requiring at least 4 counts/sector, where spurious counts are seen to disappear. Panels 3 and 4 increase the threshold to 0.25 and 0.1, respectively, where it is seen that many of the low-count rate (yet physically meaningful) higher-energy channel fluxes are removed. Therefore, it is prudent to tune the thresholding level of $dQ/Q$ to match the particular characteristics in the event of interest. 

The $dQ/Q$ values are evaluated prior to the dead time correction, which can be used to estimate the underlying counting uncertainty in L1 or L2 data products. The $dQ/Q$ values can be computed by prior loading of L0 data, or direct reading from the L2 CDF files. Filtering on the desired data can be achieved by testing whether the element-wise $dQ/Q$ values exceed the preferred error threshold, which removes low-count measurements, including both background events and physically meaningful weak fluxes which are below the instrument sensitivity. We note that $dQ/Q$ should not be confused with flux uncertainty, which can differ substantially due to count rate boosting from dead time correction (especially in the lowest energy channels).

\subsection{Example scientific applications}
Lastly, to motivate how EPD data can be used for scientific research, we provide two example published results that contributed to the closure of the ELFIN mission's primary science objectives. These include \cite{Angelopoulos2023}, which focused on EMIC wave studies, and \cite{Wilkins2023}, which focused on charged particle isotropy boundaries (IB).

\begin{figure}[!htb]
    \centering
    \includegraphics[width=\linewidth]{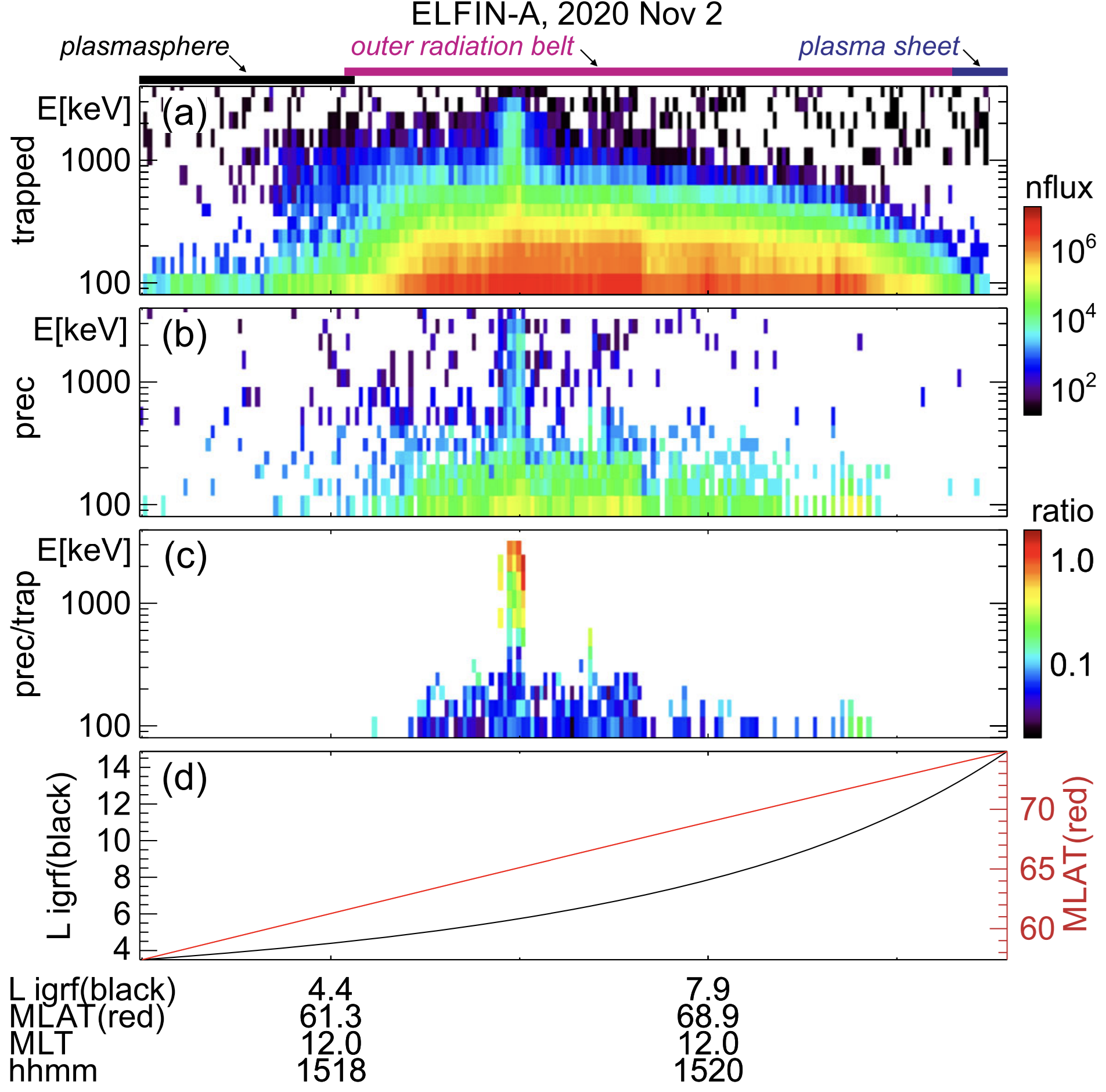}
    \caption{Example scientific application showcasing the identification of EMIC wave signatures in ELFIN EPD particle data. Modified from Fig. 6 of \cite{Angelopoulos2023}.}
    \label{fig:sci_angelopoulos2023}
\end{figure}

Fig. \ref{fig:sci_angelopoulos2023} (adapted from Fig. 6 of \cite{Angelopoulos2023}) shows an example of an electron spectrum consistent with EMIC wave scattering, based on the ratio of precipitating to perp flux at characteristic energies. Such an analysis procedure can be applied, for example to a range of events in the ELFIN database to compile statistical results, as well as investigate individual events of interest. It can also be applied to other types of waves (e.g. whistlers, ULF, etc.), and combined via conjunctions between ELFIN-A, ELFIN-B, and other satellites and ground stations to infer multi-scale phenomenological evolution in time and space.

\begin{figure}[!htb]
    \centering
    \includegraphics[width=\linewidth]{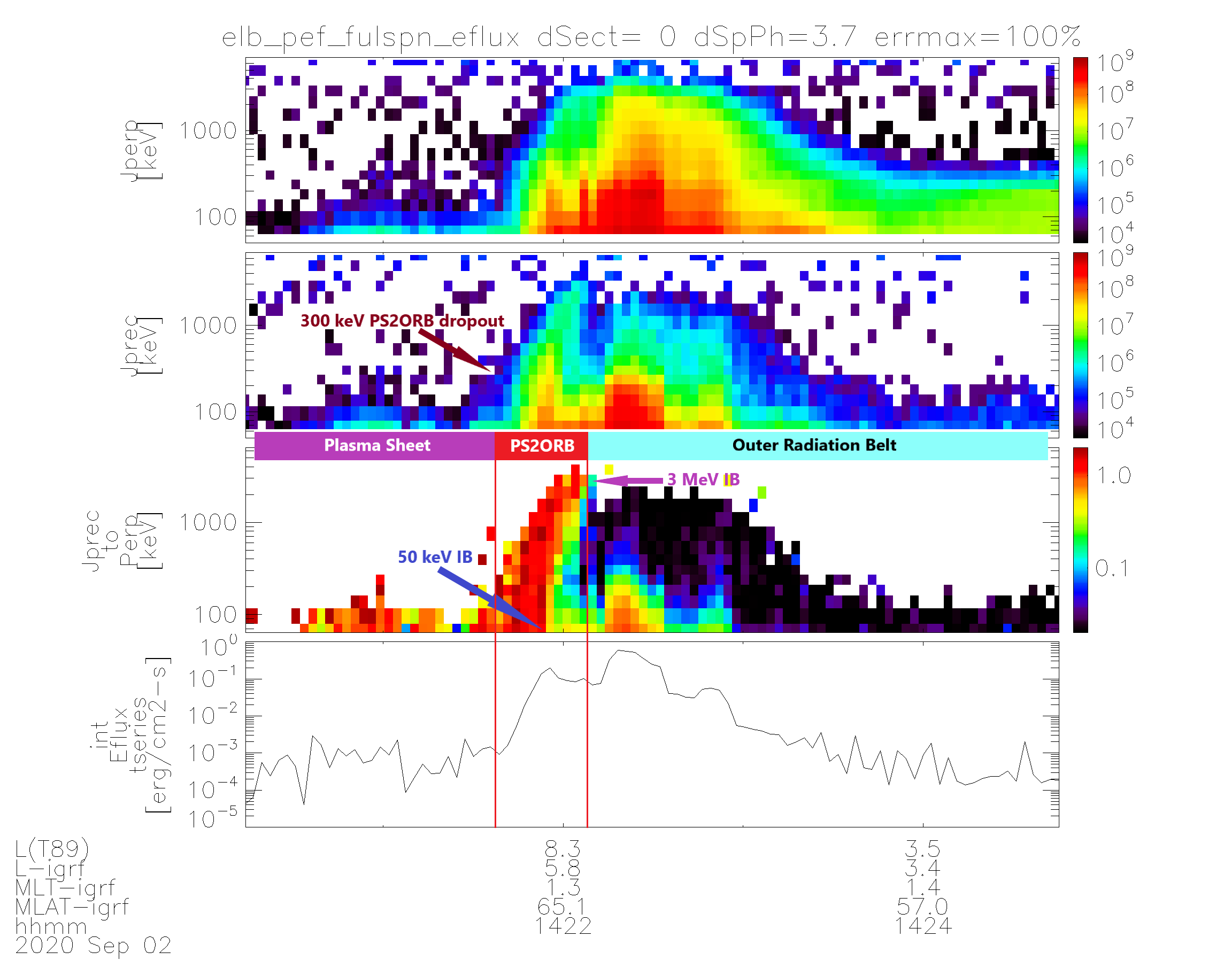}
    \caption{Example scientific application showcasing the identification of an isotropy boundary energy-latitude dispersion region. Modified from Fig. 2 of \cite{Wilkins2023}.}
    \label{fig:sci_wilkins2023}
\end{figure}

Similarly, Fig. \ref{fig:sci_wilkins2023} (adapted from Fig. 2 of \cite{Wilkins2023}) shows an example of a classical electron isotropy boundary, exhibiting a clear energy-latitude dispersion pattern consistent with field-line curvature scattering. The pattern is again identified by examining the ratio of precipitating to perp flux, and can be used to infer, for example, equatorial magnetic field characteristics deep into the magnetosphere, as well as the total amount of precipitating power from such electrons---clearly demonstrating the scientific versatility of the EPD instrument.

% \section*{Optional section: Lessons learned}
% - Optional section. Do we want to include it?

\newpage

\section*{Figures}\label{figures}

\begin{table}[!htb]
    \centering
    \begin{tabular}{|p{0.35\textwidth}|p{0.2\textwidth}|p{0.45\textwidth}|}
    \hline
        {\bf Mission (L1) Requirement} & {\bf Driving objective} & {\bf Derived EPD requirement}\\
    \hline
    \hline
        Resolve loss-cone electron pitch-angle spectra & Primary & Pitch-angle resolution $<$ 30$^\circ$ from a spinning platform \\
        Resolve electron energy spectra across the relativistic range & Primary & Relativistic electrons 0.5–4 MeV (baseline extended to 0.1 MeV); $\Delta$E/E $\lesssim$ 50\%, 16 channels\\
        Sample trapped and precipitating & Primary & Spin axis normal to orbit plane; full pitch-angle sweep twice per spin, sampling both populations on every pass\\
        Latitudinal profiles at storm-evolution cadence & Primary & $\geq$ 1 radiation-belt crossing per 4 orbits; L $\approx$ 2.5–7\\
        Attitude knowledge for pitch-angle assignment & Primary & B-field knowledge $<$ 2$^\circ$; particle FOV $<$ 28$^\circ$; spin plane contains B to within $\pm$15$^\circ$\\
        Measure plasma-sheet ions and extend electron range & Secondary & Add EPDI ion head; extend electrons to $\sim$0.1 MeV\\
    \hline
    \end{tabular}
    \caption{Traceability from ELFIN Level-1 mission requirements to EPD instrument. The Primary objective was to quantify storm-time relativistic electron precipitation and test EMIC scattering.}
    \label{tab:L1_reqs}
\end{table}

\begin{table}[!htb]
    \centering
    \begin{tabular}{|p{0.35\textwidth}|p{0.3\textwidth}|p{0.35\textwidth}|}
    \hline
        {\bf Head quantity} & {\bf Requirement} & {\bf Demonstrated capability}\\
    \hline
    \hline
        EPDE Energy range & 0.5–4 MeV & 0.05–5 MeV \\
        EPDE Energy resolution $\Delta$E/E & $<$ 50\% & Satisfied (most $<$ 40\%) \\
        EPDE Pitch-angle resolution & $<$ 30$^\circ$ & $\pm$11.25$^\circ$ (FWHM $<$ 22.5$^\circ$); 16 sectors/spin \\
        EPDE Geometric factor & $>$ 0.1 cm$^{2}$ sr & $>$ 0.13 cm$^{2}$ sr\\
        EPDE Number flux & 5$\times$10$^{5}$ cm$^{-2}$ s$^{-1}$ sr$^{-1}$ & Satisfied\\
        EPDE Field of view & $<$ 28$^\circ$ & 20$^\circ$ $\times$ 20$^\circ$\\
        EPDI Energy range & 50–300 keV & 50–5000 keV\\
        EPDI Energy resolution $\Delta$E/E & $<$ 50\% & Satisfied (most $<$ 40\%)\\
        EPDI Geometric factor & $>$ 5$\times$10$^{-3}$ cm$^{2}$ sr & $\sim$0.03 cm$^{2}$ sr\\
        EPDI Number flux & 5$\times$10$^{5}$ cm$^{-2}$ s$^{-1}$ sr$^{-1}$ & Satisfied\\
        EPDI Field of view & $<$ 28$^\circ$ & 22$^\circ$ $\times$ 22$^\circ$\\
    \hline
    \end{tabular}
    \caption{EPD requirement-versus-demonstrated-capability summary (from Table 3 of \cite{Angelopoulos2020}; EPDI energy range capability updated for the in-flight calibration reported here).}
    \label{tab:inst_performance}
\end{table}

\newpage

\begin{table}[!htb]
    \centering
    \begin{tabular}{|p{0.25\textwidth}|p{0.37\textwidth}|p{0.38\textwidth}|}
    \hline
        {\bf Data product (level)} & {\bf Element names (array size)} & {\bf Description}\\
    \hline
    \hline
        Raw sector counts (L0) & \verb|el[x]_p[s]f_raw| (16) & Raw particle impact counts per spin sector in each energy channel \\
        Sector count rates (L0) & \verb|el[x]_p[s]f_cps| (16) & Particle counts per second (cps) in each spin sector and energy channel \\
         & \verb|el[x]_p[s]f_spinper| (1) & Instantaneous satellite spin period (s) from onboard spin sectoring algorithm\\ 
        Number flux (L1) & \verb|el[x]_p[s]f_nflux| (16) & Calibrated number flux (``nflux'') in 1/cm$^2$ s str MeV in each energy channel at spin sector time resolution \\
        Energy flux (L1) & \verb|el[x]_p[s]f_eflux| (16) & Calibrated energy flux (``eflux'') in keV/cm$^2$ s str MeV in each energy channel at spin sector time resolution \\
        Pitch-angle spectra (L2) & \verb|el[x]_p[s]f_pa_[p]_[t]_omni| (16) & Calibrated omnidirectional flux (nflux or eflux) in each energy channel at spin period time resolution \\
         & \verb|el[x]_p[s]f_pa_[p]_[t]_perp| (16) & Calibrated perpendicular flux (nflux or eflux) in each energy channel at spin period time resolution \\
         & \verb|el[x]_p[s]f_pa_[p]_[t]_prec| (16) & Calibrated precipitating flux (nflux or eflux) in each energy channel at spin period time resolution \\
         & \verb|el[x]_p[s]f_pa_[p]_[t]_back| (16) & Calibrated back-scattered flux (nflux or eflux) in each energy channel at spin period time resolution \\
         & \verb|el[x]_p[s]f_pa_[p]_[t]| (16$\times N$) & Calibrated pitch-angle spectrogram (nflux or eflux) as a function of energy\\
         & \verb|el[x]_p[s]f_pa_[p]_[t]_ch0| ($N$) & Calibrated pitch-angle spectrogram (nflux or eflux), summed over the 50-160 keV energy range (default) \\
         & \verb|el[x]_p[s]f_pa_[p]_[t]_ch1| ($N$) & Calibrated pitch-angle spectrogram (nflux or eflux), summed over the 160-345 keV energy range (default) \\
         & \verb|el[x]_p[s]f_pa_[p]_[t]_ch2| ($N$) & Calibrated pitch-angle spectrogram (nflux or eflux), summed over the 345-900 keV energy range (default) \\
         & \verb|el[x]_p[s]f_pa_[p]_[t]_ch3| ($N$) & Calibrated pitch-angle spectrogram (nflux or eflux), summed over the $>$900 keV energy range (default) \\
         & \verb|el[x]_p[s]f_pa_[p]_[t]_[v]_err| & Normalized error ``$dQ/Q$'' (counting uncertainty) in particular pitch-angle product with name designation [v] \\
         & \verb|el[x]_p[s]f_pa_vals| (1) & Instantaneous pitch-angle ($^\circ$) in each spin sector \\
         & \verb|el[x]_p[s]f_losscone| (1) & Local bounce loss-cone pitch-angle ($^\circ$; 100 km terminal altitude) \\
         & \verb|el[x]_p[s]f_antilosscone| (1) & Local bounce anti-loss-cone pitch-angle ($^\circ$; 100 km conjugate altitude) \\
         & \verb|el[x]_epd_phase_delays| (1) & Best-fit pitch-angle spin phase correction ($^\circ$) applied to the data \\
    \hline
    \end{tabular}
    \caption{Selected EPD data products/levels, associated timeseries variable elements/array sizes per timeseries sample, and descriptions. The string [x] represents a placeholder for the particular satellite designation (a or b); [s] a placeholder for the particular particle species (e or i); [p] a placeholder for fullspin or halfspin pitch-angles; [t] for type of ``spec2plot'' flux data (nflux or eflux); [v] can be any pitch-angle-based product; and $N$ is an integer depending on the choice of full-spin or half-spin data. Note that the exact variable names differ between IDL SPEDAS and PySPEDAS; they can be found using the tplot\_names routine upon data loading.}
    \label{tab:data_products}
\end{table}

\newpage

\newpage

\newpage

\newpage

\newpage

\newpage

\newpage

\newpage

\newpage

\newpage

\newpage

\newpage

\newpage

\newpage

\newpage

\newpage

\newpage

\newpage

\newpage

\newpage

\newpage

\newpage

\newpage

\newpage

\newpage

\newpage

\newpage

\backmatter

% \bmhead{Supplementary information}

% If your article has accompanying supplementary file/s please state so here. 

% Authors reporting data from electrophoretic gels and blots should supply the full unprocessed scans for key as part of their Supplementary information. This may be requested by the editorial team/s if it is missing.

% Please refer to Journal-level guidance for any specific requirements.

\bmhead{Acknowledgements}
We acknowledge support by NASA award \# NNX14AN68G (2014-2023) and NSF award \# AGS-1242918 (2014-2019) and AGS-2019950 (2020-2023). We are grateful to NASA’s CubeSat Launch Initiative program for successfully launching the ELFIN satellites in the desired orbits under ELaNa XVIII. We thank the AFOSR for early support of the ELFIN program under its University Nanosat Program, UNP-8 project, contract \# FA9453-12-D-0285 (2013-2015).

\newpage
\bibliography{sn-bibliography}% common bib file
%% if required, the content of .bbl file can be included here once bbl is generated
%%\input sn-article.bbl

\end{document}